\documentclass[twocolumn]{aastex631}

\newcommand{\A}{\text{\normalfont\AA}}

\def \h2{{\rm H_{2}}}

\def \dn4000{D_{{\rm n}}(4000) }

\begin{document}

\title{What Are Those Tiny Things?\\
A First Study of Compact Star Clusters in the SMACS0723 Field with JWST}

\correspondingauthor{Andreas L. Faisst}
\email{afaisst@ipac.caltech.edu}

\author[0000-0002-9382-9832]{Andreas L. Faisst}
\affiliation{Caltech/IPAC, MS314-6, 1200 E. California Blvd. Pasadena, CA 91125, USA}

\author[0000-0001-7583-0621]{Ranga Ram Chary}
\affiliation{Caltech/IPAC, MS314-6, 1200 E. California Blvd. Pasadena, CA 91125, USA}

\author[0000-0003-2680-005X]{Gabriel Brammer}
\affil{Cosmic Dawn Center (DAWN), Copenhagen, Denmark}
\affil{Niels Bohr Institute, University of Copenhagen, Jagtvej 128, 2200 Copenhagen, Denmark}

\author[0000-0003-3631-7176]{Sune Toft}
\affil{Cosmic Dawn Center (DAWN), Copenhagen, Denmark}
\affil{Niels Bohr Institute, University of Copenhagen, Jagtvej 128, 2200 Copenhagen, Denmark}



\begin{abstract}
We use the unprecedented resolution and depth of the JWST NIRCam Early Release Observations at $1-5\,{\rm \mu m}$ to study the stellar mass, age, and metallicity of compact star clusters in the neighborhood of the host galaxies in the SMACS J0723.3-7327 galaxy cluster field at $z = 0.39$.
The measured colors of these star clusters show a similar distribution as quiescent galaxies at the same redshift, but are $>3$ magnitudes fainter than the current depths of wide-field galaxy survey. The star clusters are unresolved in the NIRCam/F150W data suggesting sizes smaller than $50\,{\rm pc}$. This is significantly smaller than star forming clumps or dwarf galaxies in local galaxies.
From fitting their photometry with simple stellar population (SSP) models, we find stellar metallicities consistent with $0.2-0.3\,{\rm Z_{\odot}}$ and ages of $1.5^{+0.5}_{-0.5}\,{\rm Gyrs}$.
We rule out metallicities $<0.2\,{\rm Z_{\odot}}$ and solar/super-solar at $4\sigma$ significance.
Assuming mass-to-light ratios obtained from the best-fit SSPs, we estimate stellar masses of $2.4^{+3.0}_{-1.5}\times 10^6\,{\rm M_{\odot}}$. These are between average masses of local globular clusters and dwarf galaxies.
Our analysis suggests middle-aged globulars with relatively recent formation times at $z=0.5-0.7$, which could have been subsequently stripped away from their host galaxies due to interactions in the cluster environment, or formed in cold flows onto the cluster core. However, we cannot rule out these objects being compact cores of stripped dwarf galaxies.
\end{abstract}



\section{Introduction} \label{sec:intro}

The Early Release Observations \citep[ERO; ][]{PONTOPPIDAN22} have already showcased the capabilities of the new {\it James Webb Space Telescope} \citep[JWST;][]{RIGBY22} in terms of its sensitivity and resolution at near-infrared wavelengths.

\begin{figure*}[t]
\centering
\includegraphics[angle=0,width=2.1\columnwidth]{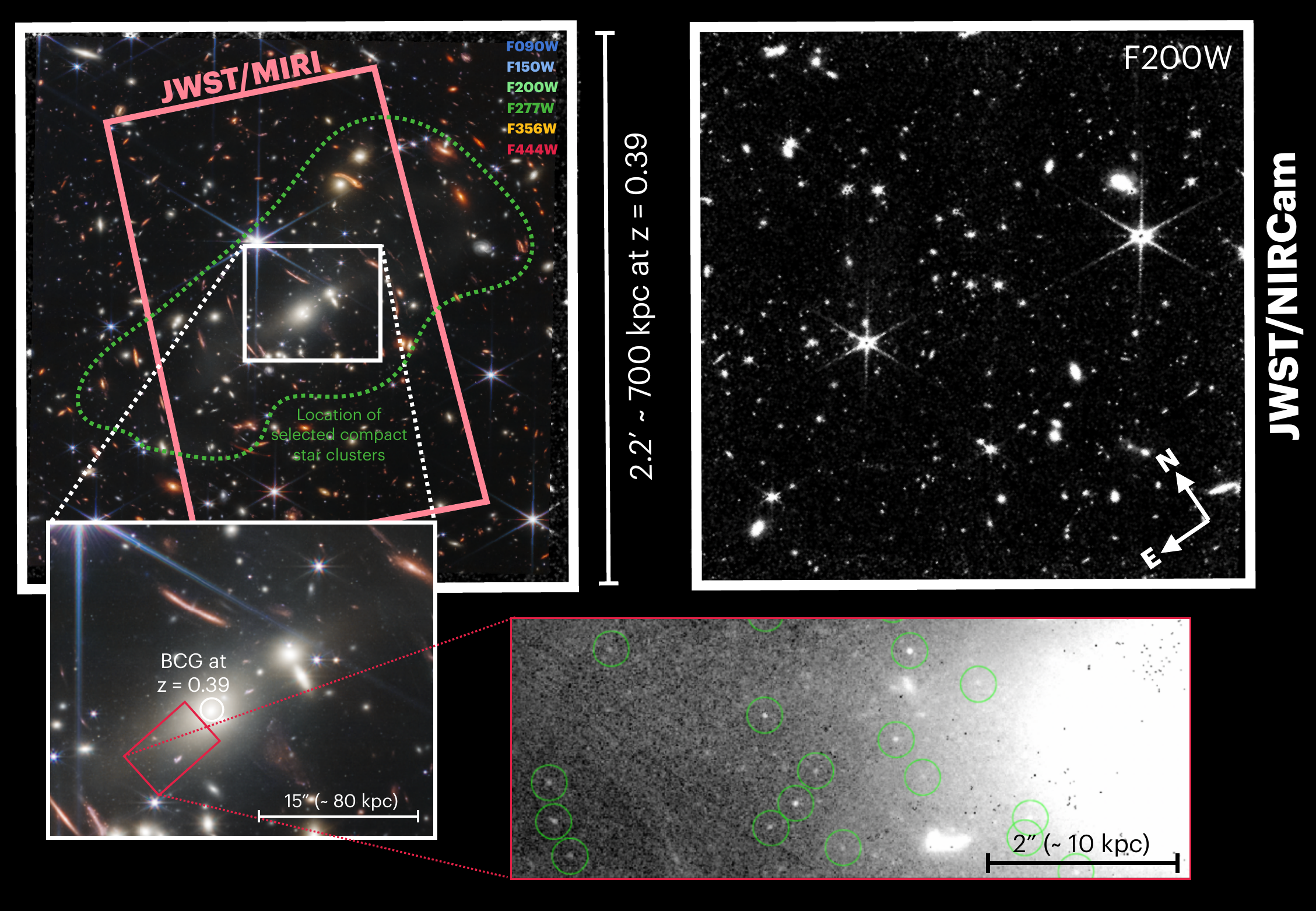}
\caption{Overview of data used in this study. The large white squares are the $2.2\arcmin \times 2.2\arcmin$ FoVs of the two NIRCam modules. The left one is centered on the cluster BCG (color composite of all $6$ observed NIRCam filters) and the right one (F200W shown only) is offset by approximately $3\arcmin$ (or $\sim1\,{\rm Mpc}$ at cluster redshift). The location of the selected compact star clusters studied in this work is marked with the green dashed contour. The JWST/MIRI coverage (red) is indicated for completeness, although MIRI data is not used in this study. The red inset shows an example selection of globulars (green circles) close to the BCG.
\label{fig:overview}}
\end{figure*}

Here, we use the unprecedented sensitivity of the JWST ERO data to study for the first time, the faintest and most compact structures around the galaxies associated with the SMACS J0723.3-7327 galaxy cluster field (hereafter SMACS0723) at $z=0.39$. Although these objects appear to be consistent with compact star clusters (or globular clusters), they could just as well be the cores of stripped dwarf galaxies.

Globulars around galaxies are abundant and have been observed in our Milky Way as well as many local galaxies \citep[e.g.,][]{HARRIS79,LARSEN01,BRODIE14,HARRIS14,MASSARI19}. The number of globulars hosted by galaxies range from a few to tens for dwarf galaxies such as the Magellanic clouds \citep[e.g.,][]{GEORGIEV10}, to a few hundreds for intermediate-mass galaxies such as our Milky Way or M31 \citep[e.g.,][]{HARRIS91,MASSARI19}, to thousands and more for massive brightest cluster galaxies (BCGs) in galaxy clusters \citep[e.g.,][]{PENG08,HARRIS17,ALAMOMARTINEZ13}, which early on were also associated with the intercluster medium \citep[ICM; e.g., ][]{JORDAN03,BASSINO03}.

There are different mechanisms suggested for the formation of globulars. Old globulars, among the oldest stellar structures in our universe \citep[e.g.,][]{RICOTTI16}, could be formed through rapid processes \citep[such as the collapse of gas clouds, e.g.,][]{PEEBLES68,FORBES18} in the early stages of the universe (such as the Epoch of Reionization at $z\sim6$). On the other hand, the peak formation of globulars (i.e., their ages) may coincide with the peak of cosmic star formation density in galaxies around $z\sim2$, suggesting they are formed via continuous star formation in gas-rich environments \citep[][]{HARRIS91,BRODIE06,FORBES18}. Indeed, some observations have suggested such a scenario by finding young globulars in dense, gas-rich merging systems \citep[][]{SCHWEIZER98,DEGRIJS01,TRUJILLOGOMEZ21} and current theoretical models mostly assume such a co-evolutionary scenario \citep[e.g.,][]{PFEFFER18,CHOKSI19,LI19,ELBADRY19}. There are also observational hints of cluster formation from cooled gas from the ICM to the galaxy cluster core \citep[e.g.,][]{HOLTZMAN92}. For an in-depth discussion of the different globular cluster formation mechanisms, we refer the readers to \citet[][]{KRUIJSSEN15}.

Age determination based on main-sequence turnoff from colors of single stars and cooling curves from long-lived white dwarfs suggest a dominant population of very old globulars with a range of metallicity as well as a ``young'' branch for which metallicity is anti-correlated to age \citep[][]{HARRIS79,KRAUSS03,MARINFRANCH09,FORBES10,VANDENBERG13,LEAMAN13,FORBES15,FORBES18,KRUIJSSEN19,MASSARI19}. The latter may be associated with disruption events of dwarf galaxies and accretion ({\it in-situ} vs. {\it ex-situ} formation, see also \citealt[][]{KRUIJSSEN19}). Among the young globulars, the metal-poor population is slightly older than the metal-rich population ($12.5$ vs. $11.5\,{\rm Gyrs}$; \citealt[][]{FORBES18}), however, these differences are of low significance due to uncertainties in the age measurements. Similarly, the absolute ages of the oldest and metal poor globulars have a significant range; Alternative age measurements by \citet[][]{FORBES15} (assuming that the metallicity of globulars follows the galaxy mass$-$metallicity relation) suggest that even the oldest, metal-poor globulars form at $z<5.9$, {\it i.e.} after the reionization of the early universe. 

The interacting galaxies in the SMACS0723 galaxy cluster environment may be the place to look for evidence for the formation of compact star clusters at $z~=~0.39$ $-$ approximately $4.3\,{\rm Gyrs}$ before the observation of Milky Way or local globulars. The galaxies are clearly interacting as shown by the evident diffuse intracluster light \citep[e.g.,][]{PASCALE22}, suggestive of stars being stripped from the galaxies. The new JWST/NIRCam observations at $1-5\,{\rm \mu m}$ show a considerable amount of relatively blue and faint ($>28\,{\rm AB\,mag}$) point sources around the cluster members \citep[see also ][]{LEE22}. As shown in Section~\ref{sec:stars}, this number of faint point sources cannot be explained by Milky Way stars, nor globulars in the Milky Way, which, at sizes of $\sim10\,{\rm pc}$ would be resolved and significantly brighter. These objects are good candidates for globular clusters and their study may inform us about the formation of globulars at higher redshifts.

The large angular coverage of the two NIRCam fields of view (FoV), being offset from each other by $\sim1\,{\rm Mpc}$ at the cluster's redshift, allow also a spatial study of globulars around the brightest cluster galaxy (BCG).
If recently formed star clusters are stripped from their host galaxies due to the interactions in the galaxy cluster, we would expect these to be relatively young and metal abundant compared to those originating from an early-formation scenario. Furthermore, the size and masses of these compact star clusters may differentiate them from stripped cores of dwarf galaxies \citep[e.g.,][]{IDETA04}.

This paper is organized as follows: In Section~\ref{sec:dataandmeasurements}, we detail the data used in this work, the selection of the compact star clusters, and various measurements. In Section~\ref{sec:results}, we present the results from our analysis. We conclude in Section~\ref{sec:end}. 
Throughout this work, we assume a $\Lambda$CDM cosmology with $H_0 = 70\,{\rm km\,s^{-1}\,Mpc^{-1}}$, $\Omega_\Lambda = 0.7$, and $\Omega_{\rm m} = 0.3$. All magnitudes are given in the AB system \citep{OKE74} and stellar masses and star formation rates (SFRs) are normalized to a \citet[][]{CHABRIER03} initial mass function (IMF) unless noted differently.

\section{Data and Measurements}\label{sec:dataandmeasurements}

\subsection{Data} \label{sec:data}
In this work, we focus on compact stellar clusters in the galaxy cluster environment SMACS J0723.3-7327 at $z = 0.39$ (07$^{\rm h}$23$^{\rm m}$13.3$^{\rm s}$, -73$^{\rm d}$27$^{\rm m}$25$^{\rm s}$).
SMACS0723 was initially discovered as part of the southern extension to the ROSAT All-Sky Survey \citep[][]{VOGES99} based {\it Massive Cluster Survey} \citep[MACS; ][]{EBELING01} and also detected by Planck \citep[][]{PLANCK11} through the Sunyaev-Zel'dovich effect.
It was then subsequently studied by the {\it Reionization Lensing Cluster Survey}  using the Hubble space telescope \citep[RELICS,][]{COE19}. A new study using JWST by \citet[][]{MAHLER22} places the total mass of the cluster at $M_{\rm <400\,kpc} = 3\times10^{14}\,{\rm M_{\odot}}$ (compared to the measurement by Planck of $M_{\rm 500} = 8\times10^{14}\,{\rm M_{\odot}}$). The BCG has a stellar mass of $\sim2 \times 10^{11}\,{\rm M_\odot}$ (close to the ``knee'' of the galaxy stellar mass function, e.g., \citealt[][]{DAVIDZON17}) and resembles a relatively featureless elliptical galaxy with little to no ongoing star formation.
Here, we use new JWST observations of the cluster \citep[ERO program ID 2736;][]{PONTOPPIDAN22} taken with the Near Infrared Camera \citep[NIRCam;][]{RIEKE05} broad-band filters F090W ($0.90\rm \mu m$), F150W ($1.50\rm \mu m$), F200W ($1.99\rm \mu m$), F277W ($2.76\rm \mu m$), F356W ($3.57\rm \mu m$), and F444W ($4.41\rm \mu m$). These filters correspond to wavelengths in the cluster's rest-frame of $0.65\rm \mu m$, $1.08\rm \mu m$, $1.43\rm \mu m$, $1.98\rm \mu m$, $2.57\rm \mu m$, and $3.17\rm \mu m$, respectively. NIRCam consists of two modules, each observing a $2.2\arcmin\times2.2\arcmin$ FoV, one centered on the cluster's BCG and the other one centered $\sim3\arcmin$ ($\sim1\,{\rm Mpc}$ at the cluster's redshift) to the south-west. The data were downloaded from the \textit{Mikulski Archive of Space Telescopes} (MAST)\footnote{\url{https://mast.stsci.edu/}} and subsequently reduced with the JWST data reduction pipeline. The reduced image data were then combined using the grism redshift and line analysis software for space-based spectroscopy \citep[\textsc{Grizli};][]{BRAMMER21} package to a final pixel scale of $0.02\arcsec/{\rm px}$ for F090W, F150W, and F200W and $0.04\arcsec/{\rm px}$ for F277W, F356W, and F444W, respectively \citep[][]{BRAMMER22_SMACSreduct}. Note that the recent \texttt{jwst\_0942.pmap} photometric calibration reference file was used with modifications as of September 5 2022 as described in the \textsc{Grizli} Github repository\footnote{\url{https://github.com/gbrammer/grizli/pull/107}}. The details of the data reduction will be presented in detail in an upcoming paper (Brammer et al., in prep).
The field was also imaged by JWST/MIRI at mid-infrared wavelengths. However, the MIRI data is too shallow for this study and is therefore not being used. 
Figure~\ref{fig:overview} shows an overview of the data used.

\begin{figure*}[t!]
\centering
\includegraphics[angle=0,width=1.2\columnwidth]{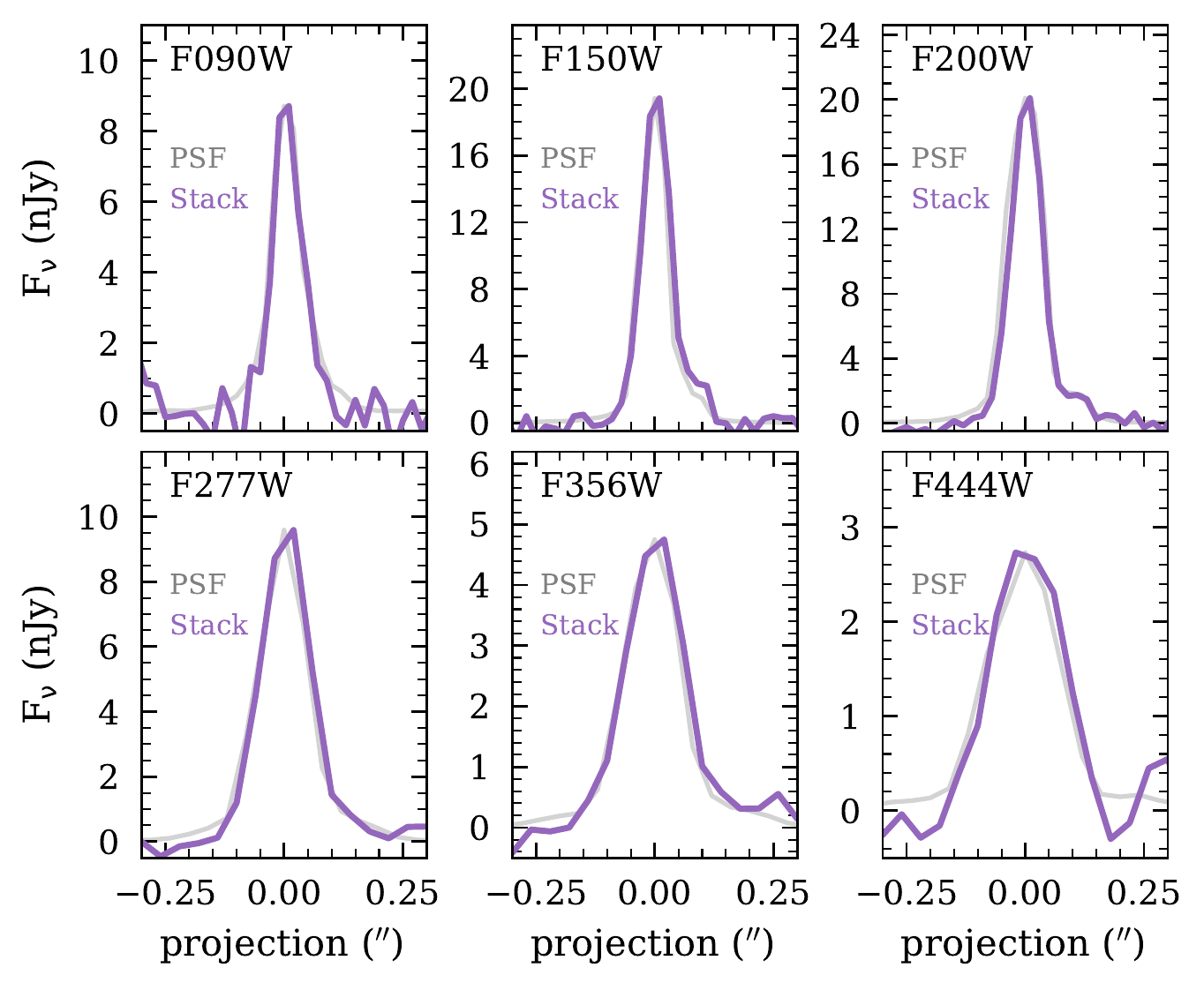}
\includegraphics[angle=0,width=0.8\columnwidth]{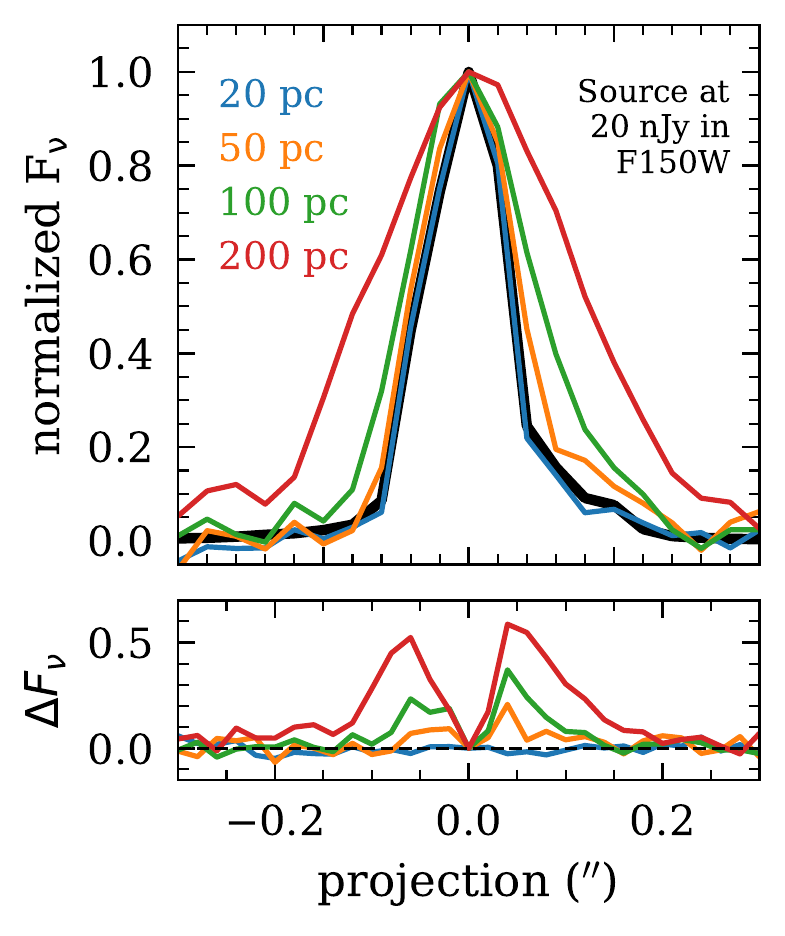}
\caption{{\it Left:} 1-dimensional light profiles of the extracted compact star clusters (purple; stacked in each filter) and the PSF (gray) in all $6$ NIRCam filters. This shows the unresolved nature of the extracted sources. {\it Right:} Simulation to assess the maximum upper limit in size of an unresolved source given the PSF in the NIRCam/F150W filter (black). We simulated sources with half-light radii of 20, 50, 100, and $200\,{\rm pc}$ (blue, orange, green, and red lines) at $z=0.39$ and a peak flux of $20\,{\rm nJy}$, consistent with observations of the F150W stack (see left panel). We convolved those profiles with the PSF (which is shown as the thick black line) and include realistic noise from the real observations. 
We find that we would be able to resolve sources larger than $50\,{\rm pc}$. Hence, the selected compact star clusters must have sizes less than $50\,{\rm pc}$. The lower panel shows the residual (source$-$PSF profile).  
\label{fig:psfs}}
\end{figure*}

\subsection{Source Selection} \label{sec:selection}
In this work, we study the faintest, compact stellar clusters in the SMACS0723 field, such as potential globular clusters around the cluster host galaxies. These sources are selected manually mainly on the F200W image, however, we require them also to be detected in at least two other bands in order to reject potential spurious sources. In addition, the sources are selected to be unresolved point sources as well as faint (to exclude Milky Way stars, see Section~\ref{sec:stars}). For example, a good upper limit in size of globulars is $\sim10\,{\rm pc}$ \citep[e.g.,][]{LARSEN01}, which, at $z=0.39$, corresponds to an angular size of $\sim1.4$ milliarcseconds (note that there are some larger globulars, for example NGC 2419 in the Milky Way and Lindsay 1 in the Small Magellanic Cloud). This is significantly smaller than the PSF size (between $20$ and $120$ milliarcseconds at $1-5\,{\rm \mu m}$), thus motivating the point-source assumption. Milky Way stars are the most likely contaminants in this selection, as discussed in Section~\ref{sec:stars}.
We tried an automated selection of globulars, however, we found that this approach results in a severely incomplete selection due to blending with the BCG and other cluster galaxies, as well as the variation of the intracluster and scattered light across the field.
In total, we identified $178$ compact star clusters in the full FoV of NIRCam around the cluster host galaxies (see Figure~\ref{fig:overview}), which we characterize and study in the following. The coordinates and NIRCam fluxes of the extracted star clusters are listed in Table~\ref{tab:photo}.
We note that our selection is not complete, but serves as a starting point to understand the properties and origin of these faint sources. We refer to the work by \citet[][]{LEE22} for a detailed study of the distribution and number density of these star clusters in the ICM of SMACS0723. If these sources are globular clusters, we would expect to only see the tip of the iceberg ({\it i.e.} the brightest ones) even at the unprecedented depths of these observations, assuming the commonly used Gaussian globular cluster luminosity function \citep[e.g.,][]{ALAMOMARTINEZ13}. 

\subsection{Measurement of the PSF}\label{sec:psf}
To measure the photometry of unresolved point sources, a robust determination of the point spread function (PSF) is crucial. Here, we measure the PSF in all $6$ NIRCam filters by stacking stars on the full FoV of the available observations. For the selection of stars, we use an identical approach as in \citet[][]{FAISST22}. Summarizing, the stars are selected based on the $R_{e}$ vs. magnitude diagram (produced by the \texttt{flux\_radius} and \texttt{flux\_auto} quantities derived by \textsc{SExtractor} \citep{Bertin1996} run the NIRCam images). We note that, based on the upturn in size measurements on that diagram, point sources at $\sim25\,{\rm mag}$ already start to enter the non-linear regime towards saturation at the depth of these NIRCam observations. We therefore require stars between $25\,{\rm mag}$ and $27\,{\rm mag}$, to avoid the aforementioned saturation effect as well as to exclude potential faint, spurious sources. In addition, a half-light radius of less than $3\,{\rm pixels}$ is required to select unresolved sources. 
For each of the $58$ stars, we create a $2\arcsec \times 2\arcsec$ cutout and subsequently center the stars before stacking them to obtain the final PSF. The final PSF for each filter (normalized and of the same pixel scale as the images, see Section~\ref{sec:data}) are available for download\footnote{\url{https://github.com/afaisst/JWST_SMACS_PSFs}}.

The relatively low number density of stars does not allow us to quantify variations of the PSF. However, the recent study by \citet[][]{NARDIELLO22} measured PSF variations across the NIRCam FoV based on stars in the globular cluster M 92, finding variations in the FWHM of up to $15-20\%$. Such variations have a negligible effect on the measured photometry in our case, but we will include this effect later when we constrain the sizes of our compact star clusters.
\vspace{-0.1cm}

\subsection{Measurement of Photometry}\label{sec:photomeasure}

The photometry of the selected globulars is measured using the software \textsc{Tractor}\footnote{\url{http://thetractor.org/}} \citep{LANG16b,LANG16a,WEAVER22b}, which performs a prior-based forced photometry including the PSF of the observations.
We first compared the 1-dimensional projected light profiles from the stack of all compact star clusters to the one of the PSF (left panel of Figure~\ref{fig:psfs}). This test shows that the star clusters are indeed consistent with unresolved point sources, hence we use the point-source fitting option of \textsc{Tractor}.
In the following, we fit the photometry of individual sources to study the variations in the stellar population across our sample. We also derive the photometry from the stacks of the sources in each of the $6$ NIRCam filters to study the average properties. In both cases, we create cutouts of $0.8\arcsec$ size of all the sources. The stacks were created by first performing a local background subtraction (measured in a sky annulus with inner radius of $5\,\rm px$ and outer radius of $10\,\rm px$) on the individual cutouts and then combining them via median stacking. The following steps are then identical for the fitting of individual sources and the stacks.
\textsc{Tractor} is run on the $0.8\arcsec$ cutouts by applying a point source model, using the F200W position of the sources as prior (determined from the 0-order moment of light on the source). During the fitting, the position is free to vary within $\pm2\,{\rm pixels}$ of the prior position. This helps to mitigate potential misalignments between the different image products.  In addition, we let \textsc{Tractor} fit and subtract the (slowly varying on $0.8\arcsec$ scales) local background at the position of the source to remove the contribution of the halo from nearby cluster galaxies and the intracluster light. Note that for that reason, simple aperture photometry will result in overestimated flux measurements. The inverse variance image (used for error estimation) is created as $1/\sigma^2$ with $\sigma$ derived from $\sigma-$clipping statistics on a large region of the image not contaminated by the intracluster light and other galaxies. The variance output by \textsc{Tractor} is used to obtain the uncertainties on the flux measurements.

\begin{figure}[t]
\centering
\includegraphics[angle=0,width=\columnwidth]{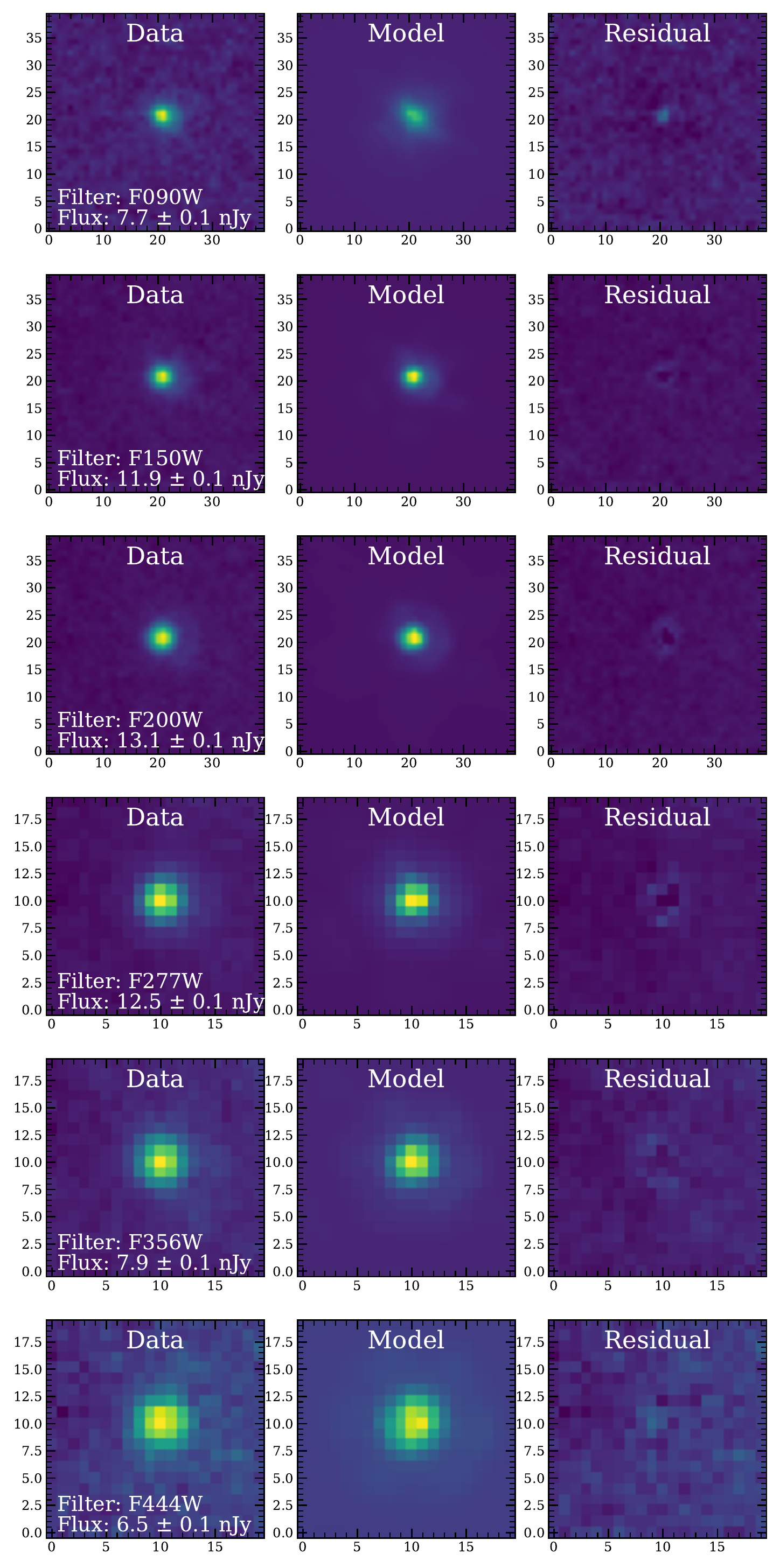}
\caption{We use the \textsc{Tractor} software package to fit point source profiles to the selected compact star clusters in the highly blended cluster environment. The first column shows the stacked cutouts in each of the NIRCAM band. The second and third columns show the best-fit model and the residual, respectively. The rows show the fits for the $6$ NIRCam filters. The measured total flux values in nJy are indicated. The intensity scale is the same for each row (the residuals are consistent with zero) and all cutouts are $0.8\arcsec$ in size.
\label{fig:fitexample}}\vspace{1.2cm}
\end{figure}

Figure~\ref{fig:fitexample} shows the fits to the stacked cutouts of the compact star clusters in the six different bands. The point source assumption is also validated by the residuals, which are consistent with zero.
The NIRCam $6$-band photometry derived for the stack is listed in Table~\ref{tab:photostack}.\vspace{+0.7cm}

\begin{figure*}[t]
\centering
\includegraphics[angle=0,width=2.1\columnwidth]{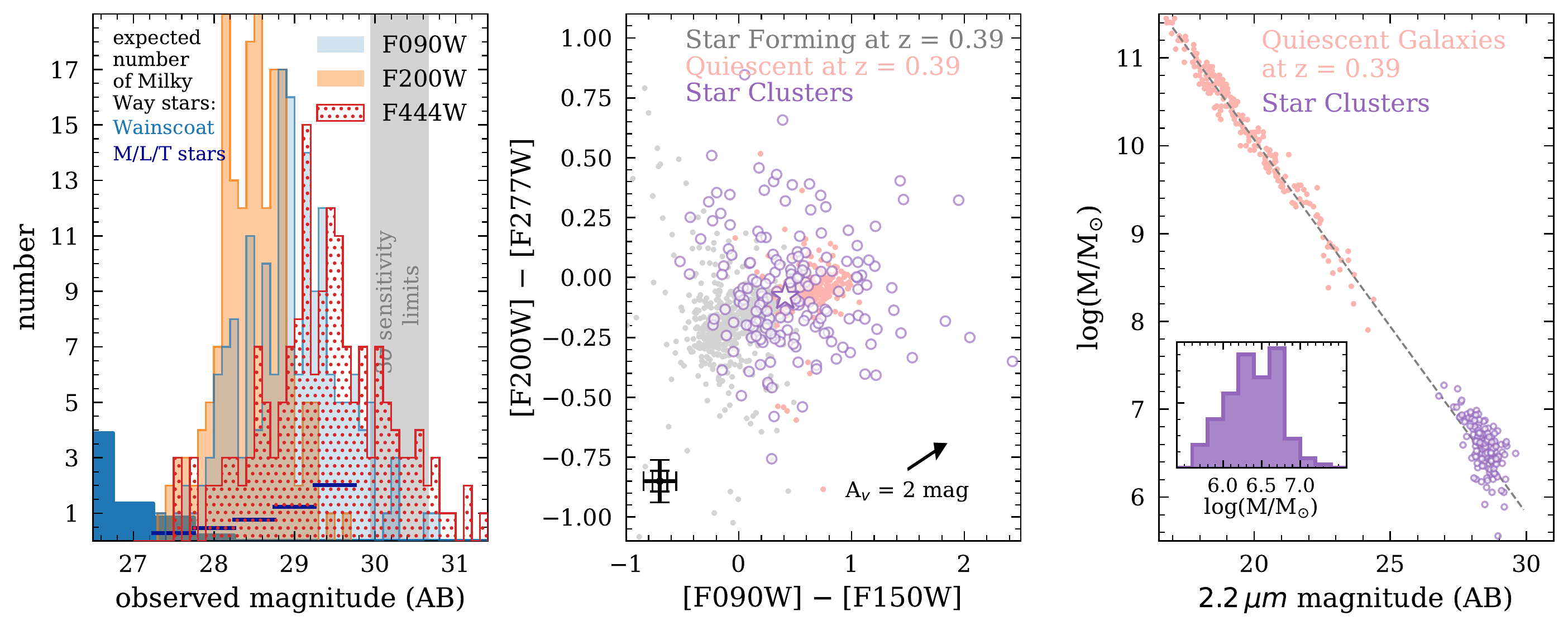}
\caption{Properties of the compact star clusters found in the SMACS0723 field.
{\it Left panel:} Magnitude distribution in F090W (blue), F200W (orange), and F444W (red semi-filled). The vertical gray area shows the range in $3\sigma$ limits in the three filters shown. The blue filled histogram shows the expected number of Milky Way stars from \citet[][]{WAINSCOAT92} towards the cluster field within the full NIRCam FoV. The dark blue horizontal lines show the same for the combined number of M/L/T stars from \citet[][]{KIRKPATRICK21}.
{\it Middle panel:} The distribution of the extracted compact star clusters (purple circles; median as purple star) in color space. Typical errors in the colors are shown by the error bars in black (lowest $10\%$ and median errors). The distribution of (de-reddened) colors of star forming (gray) and quiescent (red) galaxies at $z=0.39$ from the COSMOS2020 catalog \citep[][]{WEAVER22a} is shown. The star clusters match with the color distribution of the quiescent galaxy population. The A$_{\rm  v}=2\,{\rm mag}$ extinction vector is also shown. 
{\it Right panel:} The $2.2\,{\rm \mu m}$ magnitude vs. stellar mass relation of quiescent galaxies (red dots) is similar to the one of the star clusters, indicating a similar $K-$band M/L ratio. The stellar masses of the star clusters (median of $2.4^{+3.0}_{-1.5}\times 10^6\,{\rm M_{\odot}}$, see inset) are derived from the M/L ratio of the best-fit SSP models (Section~\ref{sec:masses}).
\label{fig:properties}}
\end{figure*}
 
\subsection{SED Fitting}\label{sec:sedfitting}

To study the properties of the compact star clusters, we fit simple stellar population (SSP) models to the photometry of the median stacks obtained in Section~\ref{sec:photomeasure}. We assume that the star clusters have been formed in a single burst, thus are well described with an SSP model. Given the low sampling in wavelength of these data, we think that this is a good approximation as multiple stellar populations would be difficult to tell apart if they exist.
Before fitting, we need to correct the extracted flux densities for Milky Way dust extinction, which we obtain from the \citet[][]{SCHLEGEL98} dust maps\footnote{Specifically, we use the \texttt{sfdmap} Python package (\url{https://github.com/kbarbary/sfdmap})} recalibrated by \citet[][]{SCHLAFLY11}. The dust extinction towards the SMACS0723 is significant at an $E(B-V)\sim0.2\,{\rm mag}$ level, which results in an wavelength-dependent extinction of $\sim 0.25\,{\rm mag}$ at $0.9\,{\rm \mu m}$, $\sim 0.07\,{\rm mag}$ at $2.0\,{\rm \mu m}$, and less than $0.03\,{\rm mag}$ at $4.4\,{\rm \mu m}$ assuming the \citet[][]{FITZPATRICK99} dust extinction curve. We use these values to correct the measured flux densities of the star clusters.
For our fiducial SSP models, we chose Padova 1994 stellar evolutionary tracks \citep[][]{BERTELLI94} together with the Basel Stellar Library \citep[BaSeL, version 3.1;][]{LEJEUNE97,LEJEUNE98,WESTERA03} complemented by the empirical STELIB library \citep[][]{LEBORGNE03}\footnote{\url{http://svocats.cab.inta-csic.es/stelib/index.php}} at wavelengths blueward of $9500\,{\rm \AA}$ \citep[see][]{BRUZUALCHARLOT03}. In the following, we assume a \citet{CHABRIER03} IMF.
These fiducial SSP models were created using the software \textsc{GALAXEV}\footnote{\url{http://www.bruzual.org/bc03/}} in a grid of different ages and stellar metallicities. Specifically, the age grid ranges from $0.5\,{\rm Gyrs}$ to $11\,{\rm Gyrs}$ in steps of $0.5\,{\rm Gyrs}$ and we use $6$ different stellar metallicities of $0.005$, $0.02$, $0.2$, $0.4$, $1$, and $2.5\,{\rm Z_{\odot}}$.
We compared the results using these fiducial models to other SSP models as well. Specifically, we compared to the MIST stellar tracks \citep[][]{DOTTER16,CHOI16}, based on the MESA isochrones \citep{PAXTON11,PAXTON13,PAXTON15,PAXTON18}\footnote{\url{https://waps.cfa.harvard.edu/MIST/index.html}}, assuming a BaSeL stellar library. These additional SSP models are created using the {\it Flexible Stellar Population Synthesis} code \citep[FSPS;][]{CONROY09,CONROY10}\footnote{\url{https://github.com/cconroy20/fsps}. A Python wrapper \citep[][]{JOHNSONFSPSPython21} exists at this link: \url{https://dfm.io/python-fsps/current/}.} for a Chabrier IMF and using the same age and metallicity grid. Generally, we find that wavelengths blueward of $\sim1\,{\rm \mu m}$ rest-frame are most affected by the choice of these different models. We will comment on the differences in the results in Section~\ref{sec:results}.
The fit to the various models was performed by $\chi^2$ minimization using the \texttt{SciPy} {\it least-squares} \textsc{Python} package.
We note that we fixed the redshift of the compact star clusters to the one of the galaxies in the SMACS0723 cluster ($z=0.39$). In Section~\ref{sec:photoandcolors} we will show that this is indeed a good assumption based on the similarity in colors of the star clusters and galaxies at $z=0.39$. Also, small changes in redshift (e.g., due to the kinematics in the cluster environment) do not have an impact on the fitted properties.

\begin{figure*}[t]
\centering
\includegraphics[angle=0,width=2.1\columnwidth]{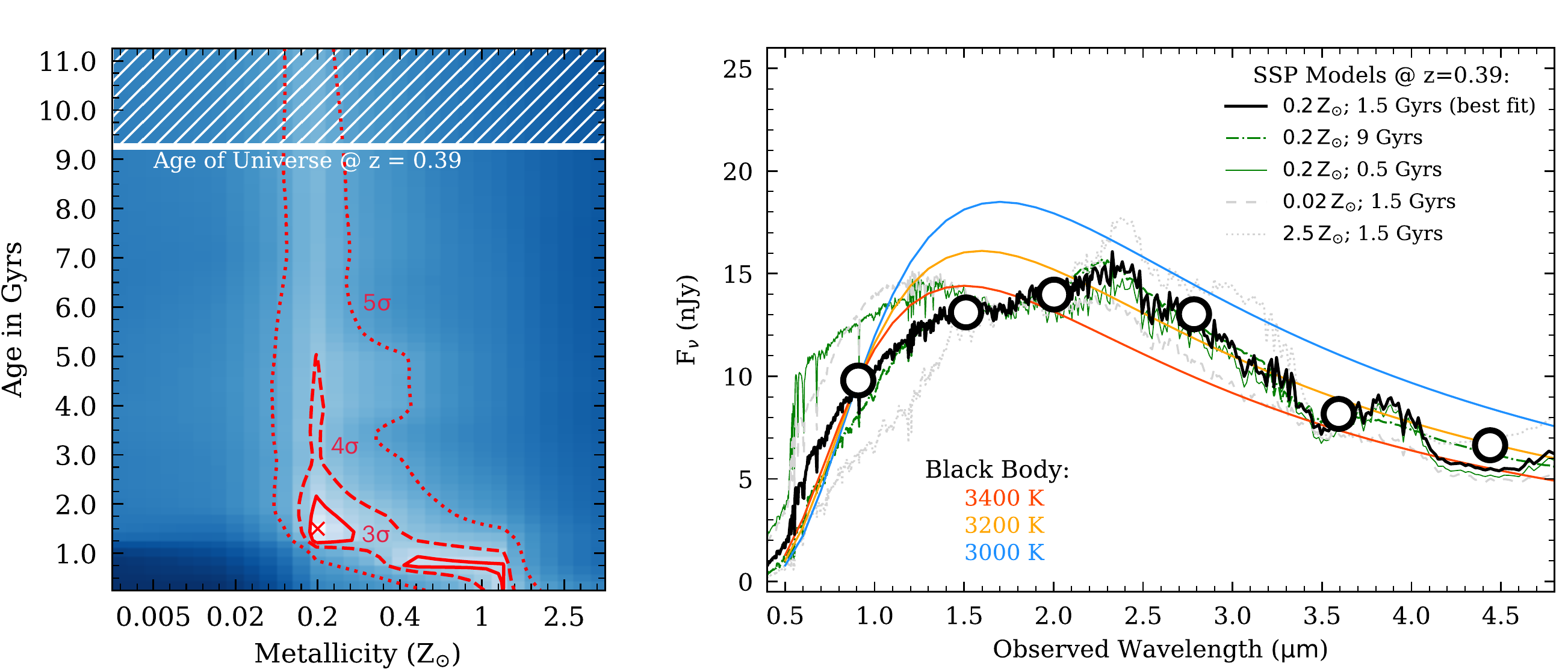}
\caption{Results from fitting the photometry of all extracted compact star clusters.
{\it Left Panel:} $\chi^2$ map of the age vs. metallicity grid with $3\sigma$, $4\sigma$, and $5\sigma$ contours indicated. The cross marks the best-fit solution ($0.2\,{\rm Z_{\odot}}$ and $1.5\,{\rm Gyrs}$). The age of the universe at $z=0.39$ ($9.3\,{\rm Gyrs}$) is indicated by the horizontal white line. For the full globular population we find an age of $1.5^{+0.5}_{-0.5}\,{\rm Gyrs}$ and a best-fit metallicity of $20-30\%$ solar. Metallicities of $<20\%$ or more than solar are ruled out at high significance. 
{\it Right Panel:} Data (black symbols) and best-fit SSP model (black line) shown with other models of different age and metallicity combinations (see legend). While different metal abundances produce different SEDs at $<1.5\,{\rm \mu m}$, older ages are difficult to discern. Better wavelength sampling with medium-band filters at $2-3\,{\rm \mu m}$ and $4-4.5\,{\rm \mu m}$ wavelengths would constrain older ages better. The colored lines show black body models at different temperatures normalized to $1\,{\rm \mu m}$.
\label{fig:fitresults}}\vspace{3.5mm}
\end{figure*}

\section{Results}\label{sec:results}

\subsection{Properties of Globular Clusters}\label{sec:properties}

\subsubsection{Photometry and Colors}\label{sec:photoandcolors}
Figure~\ref{fig:properties} show the observed photometric properties of the extracted compact star clusters in SMACS0723. As shown on the left panel, the brightness of the star clusters ranges from $28$ to $30\,{\rm mag}$ in F090W and $27.5$ to $29.5\,{\rm mag}$ in F200W (left panel). The NIRCam filters cover the (redshifted) $2\,{\rm \mu m}$ peak, hence the F090W and F444W brightness is similar.
The middle left panel shows the color distribution of the compact star clusters (purple circles) compared to star forming (gray) and quiescent (red) galaxies selected in the COSMOS2020 catalog \citep[][]{WEAVER22a} at a redshift $z=0.39$ with $\Delta z~=~0.2$\footnote{We interpolate the Suprime-Cam/$z$, UltraVISTA/$H$, UltraVISTA/$K_s$, and \textit{Spitzer}/IRAC $3.6\,{\rm \mu m}$ fluxes from COSMOS2020 to obtain the corresponding fluxes in the JWST/NIRCam filters F090W, F150W, F200W, and F277W.}. Due to their SED shape, which is due to the redshifting of the 1.6$\mu$m bump through the bandpasses, the globulars reside mostly at red [F090W]$-$[F150W]$\,\gtrsim0$ colors and blue [F200W]$-$[F277W]$\,\lesssim0$ colors. This parameter space is similar to quiescent galaxies at the same redshift, suggesting similar stellar populations. However, the star clusters are significantly fainter, $2-4\,{\rm mag}$ below the detection limits of the COSMOS surveys (right panel).

\subsubsection{Population-Averaged Ages and Metallicities}\label{sec:agesandmetal}

We note that these star clusters are faint even for JWST and most of the individual star clusters are detected at very low S/N in blue and red bands where their SED flux drops. We therefore first fit SSP models (see Section~\ref{sec:sedfitting}) to the photometry derived from the stacked cutouts to obtain a population-averaged median age and stellar metallicity measurement. The following values for metallicity and age are therefore an average of the studied relatively massive cluster population (see also Section~\ref{sec:localgcs}). Later in Section~\ref{sec:variations} we will fit individual star clusters to study the population variations of these properties.

Figure~\ref{fig:fitresults} shows the results from SED fitting using our fiducial SSP models (Padova stellar tracks) to the stacked photometry. The left panel shows a $\chi^2$ map for the adopted age and metallicity grid with $3\sigma$, $4\sigma$, and $5\sigma$ contours indicated. We find a best-fit age of $1.5^{+0.5}_{-0.5}\,{\rm Gyr}$ and a best-fit metallicity of $0.2-0.3\,{\rm Z_{\odot}}$ ($3\sigma$). Interestingly, the $\chi^2$ map also indicates the presence of a younger ($<1\,{\rm Gyr}$) and more metal rich ($0.4-1.0\,{\rm Z_{\odot}}$) solution. The right panel of Figure~\ref{fig:fitresults} shows the data (black symbols) with the best-fit SED (black line) as well as other models (see legend in figure) for comparison with the same age or metallicity as the best fit.

The data excludes with high significance metallicities below $0.2\,{\rm Z_{\odot}}$ as well as solar/super-solar. This is because changes in metallicity affect the colors significantly at a fixed age. For example, compare the $1.5\,{\rm Gyr}$ model at $0.02\,{\rm Z_{\odot}}$ (gray dotted line) and at $2.5\,{\rm Z_{\odot}}$ (gray dashed line) with the best fit ($0.2\,{\rm Z_{\odot}}$) in the right panel of Figure~\ref{fig:fitresults}.

On the other hand, ages are not constrained well by our current data, especially with the JWST filters available. For example, ages up to $5\,{\rm Gyrs}$ cannot be excluded at better than $4\sigma$ level and maximal ages of the Universe at $z=0.39$ ($9.3\,{\rm Gyrs}$) are possible at $5\sigma$ level. This is expected by the lack of blue coverage of the JWST filters, and in addition the relation between color and age only changes weakly for SSPs with ages older than $\sim3\,{\rm Gyr}$. For example, the [F090W]$-$[F200W] color (corresponding to rest-frame $(r-H)$ color) changes by $0.4-0.6\,{\rm mag}$ (depending on metallicity) for ages $<3\,{\rm Gyr}$ (caused by hot massive stars), while it only changes by $\sim0.2\,{\rm mag}$ thereafter.
We note that a better sampling of the observed $0.5-1.5\,{\rm \mu m}$ regime could constrain the older ages better (see Figure~\ref{fig:fitresults}).

In addition, we quantify the reliability of these results by comparing to fits using the MIST isochrones as discussed in Section~\ref{sec:sedfitting}. We find that the models differ significantly at wavelengths blueward of $\sim1\,{\rm \mu m}$ rest-frame for low metallicities and low ages. However, we find overall that our results are robust against the choice of different model parameterizations.
Specifically, using the {\it MIST} isochrones, we find consistent ages of $1.5^{+0.5}_{-0.5}\,{\rm Gyr}$ and a best-fit metallicity of $0.4\,{\rm Z_{\odot}}$ with a slightly larger $3\sigma$ range of $0.2-0.4\,{\rm Z_{\odot}}$. Similarly to the other models, metallicities of $<0.2\,{\rm Z_{\odot}}$ as well as solar metallicities are robustly excluded.\vspace{10mm}

\subsubsection{Stellar Masses}\label{sec:masses}
From the best-fit SSP models obtained for each individual star cluster, we derive a $V-$band mass-to-light (M/L) ratio from which we can estimate the total stellar masses of the compact star clusters. For the best-fit SSP model fit to the stacked photometry, we derive a M/L ratio of $0.45\,{\rm M_{\odot}\,L_{\odot}^{-1}}$. The M/L ratios of the individual star clusters range from $0.2-4.8\,{\rm M_{\odot}\,L_{\odot}^{-1}}$. With the $V$-band luminosity derived from the NIRCam/F090W filter (which is close to the $\sim5500\,{\rm \AA}$ $V$-band in rest-frame), we derive stellar masses for the individual star clusters and we find a median of $2.4^{+3.0}_{-1.5}\times 10^6\,{\rm M_{\odot}}$ (using the M/L ratio derived from the stacked SED would result in a stacked median stellar mass of $3.9^{+3.2}_{-1.8}\times 10^6\,{\rm M_{\odot}}$).

As shown on the right panel in Figure~\ref{fig:properties}, the star clusters line up well with quiescent galaxies at $z=0.39$ on the $K-$band magnitude vs. stellar mass diagram. This suggest that these compact star clusters have a similar M/L ratio as quiescent galaxies at the same redshift. Assuming the quiescent galaxy M/L relation, we would obtain an average stellar mass of $3.3^{+1.6}_{-1.1}\times 10^6\,{\rm M_{\odot}}$.

We can compare this obtained M/L ratio to what is measured for globular clusters in the Milky Way and local galaxies. The M/L ratios depend on the age (hence mass itself) and metallicity of the stellar population. Given the estimated age and metallicity of our star clusters, we find a possible range in M/L of $1-4\,{\rm M_{\odot}\,L_{\odot}^{-1}}$ in rest-frame $V$-band according to the observations and models in \citet[][]{KRUIJSSEN08}. Note that the M/L ratio increases with age, hence the upper end of this range would be applicable to old globulars, while the lower end is more common in young star clusters. Our measured M/L is consistent with younger ages.

\begin{figure}[t]
\centering
\includegraphics[angle=0,width=\columnwidth]{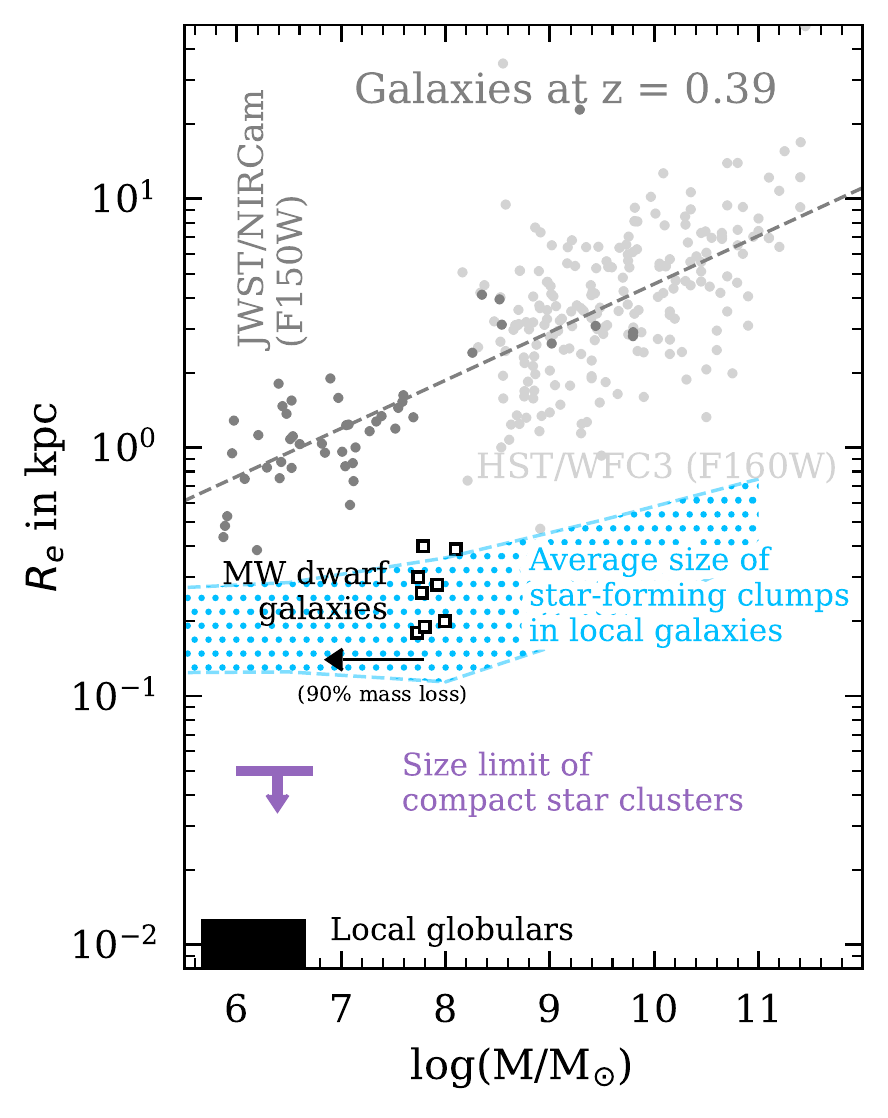}
\caption{Upper limits on the sizes of the compact star clusters (purple) compared to sizes of $z=0.39$ galaxies from HST COSMOS-DASH \citep[][]{MOWLA19,CUTLER22} in WFC3/F160W (light gray dots) and measured from NIRCam/F150W at lower masses (dark gray dots). The line shows a linear fit to the $M-R_{\rm e}$ relation. Also indicated are typical sizes of star-forming regions in local galaxies \citep[light blue hatched with $1\sigma$ percentile;][]{LARSON20}, the sizes/masses of dwarf galaxies around the Milky Way \citep[black squares;][]{STRIGARI08}, and sizes/masses of local globular clusters \citep[black box;][]{LARSEN01,HILKER20}.
The $z=0.39$ compact star clusters overlap with the massive end of the stellar mass distribution of local globulars as well as with common masses of Milky Way dwarf assuming they lose $\sim90\%$ of their mass during close passages to the BCG (black arrow).
\label{fig:sizes}}\vspace{-0cm}
\end{figure}

Summarizing, the masses derived for our compact star clusters are on the high-end of what is expected for globular star clusters in local galaxies (see also Figure~\ref{fig:sizes}). Furthermore, the high-mass end of the derived mass range may overlap with stellar masses estimated for Milky Way dwarf galaxies. To assess this, we compare the size constraints on our compact star clusters with the sizes of local dwarf galaxies.

\subsubsection{Sizes}\label{sec:sizes}
As established already in Section~\ref{sec:photomeasure} and Figure~\ref{fig:psfs}, the selected compact star clusters are unresolved, hence we are only able to place an upper limit on their sizes. A conservative upper limit would be the NIRCam PSF size at $\sim1.5\,{\rm \mu m}$, which is $\sim180\,{\rm pc}$ at $z=0.39$. To measure a more accurate upper size limit, we carry out a simple simulation as shown in the right panel of Figure~\ref{fig:psfs}. Specifically, we simulate point sources of different sizes (from $20$ to $200\,{\rm pc}$), which we convolve with the NIRCam/F150W PSF (Section~\ref{sec:psf}). We add realistic background noise measured from the real NIRCam/F150W image and assume a point source peak flux of $20\,{\rm nJy}$, which is consistent with the one measured for the stack in the same filter (see left panel of Figure~\ref{fig:psfs}). This simple simulation shows that we would be able to resolve sources larger than $50\,{\rm pc}$. We therefore place an upper limit to the sizes of our compact star clusters of $50\,{\rm pc}$.

Figure~\ref{fig:sizes} compares this size limit to other Galactic and extra-galactic sources at similar stellar masses as derived for our compact star clusters (Section~\ref{sec:masses}).
The sizes of the galaxies at $z=0.39$ are taken from the HST COSMOS-{\it DASH} morphology catalog \citep[][]{MOWLA19,CUTLER22} in HST WFC3/F160W. At lower masses, we complement that sample with galaxies at the same redshift directly extracted from the NIRSpec/F150W image\footnote{The masses and redshift are computed using the SED-fitting code \textsc{EAZY} \citep[][]{BRAMMER08,BRAMMER22_SMACSreduct}. Note, that we only extract sources from the NIRCam field offset from the cluster's BCG to avoid the effects of lensing on the size measurements.}. The sizes of these lower-mass galaxies are consistent with extrapolating the $M-R_{e}$ relation at $>10^9\,{\rm M_{\odot}}$ (dashed line) to these lower masses. The galaxies are clearly more extended than the compact star clusters selected here.
We also show average sizes of star-forming clumps (blue hatched area showing $1\sigma$) in local galaxies taken from \citet[][]{DRAZINOS13} and \citet[][]{LARSON20}. At the resolution of the NIRCam observations, we would be able to marginally resolve such structures if they were present in the (quiescent) SMACS0723 host galaxies.
Similarly, we would resolve objects similar in mass and size as the Milky Way dwarf galaxies, here taken from \citet[][]{STRIGARI08}\footnote{The masses were derived from total luminosities given in \citet[][]{STRIGARI08} and assuming the mass-to-light ratios of the galaxies measured by \citet[][]{REVAZ09}.}. We caution however, if the dwarf galaxies have been tidally stripped in the dense cluster environment such that only the bright central core remains, we would likely not be able to distinguish those from compact star clusters. For example, a dwarf galaxy could lose more than $90\%$ of its mass during the first few pericenter passages around its host galaxy \citep[][]{IDETA04}. This would bring them to a mass consistent with our compact star clusters.

\begin{figure}[t]
\centering
\includegraphics[angle=0,width=1.0\columnwidth]{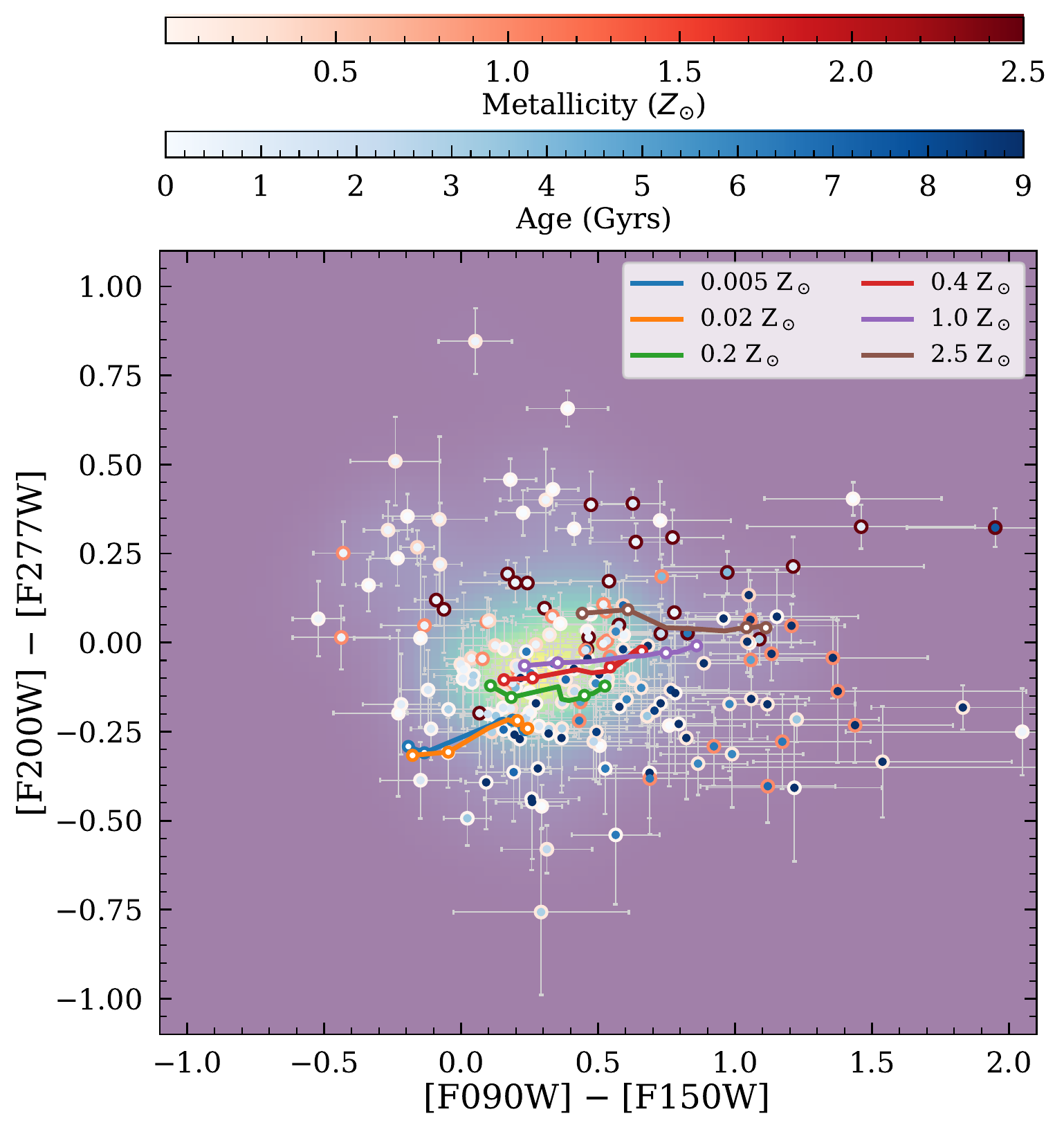}
\caption{Individual compact clusters color-coded by their best-fit metallicity (red scale) and age (blue scale). Note that each symbols has two colors assigned (metallicity as edge-color and age as face-color). Light symbols correspond to star clusters with low ages and metallicities, while darker symbols correspond older and more metal rich star clusters.
The background shows a Gaussian kernel density estimation of the data points for visual guidance of the density of points. The lines show models at different metallicities (color of lines) as a function of age (running from left to right, dots on lines represent ages of 0.5, 1, 4, and $9\,{\rm Gyrs}$).
\label{fig:variations}}
\end{figure}

\subsection{Variations in the Properties of Star Clusters}\label{sec:variations}

As shown in the middle panel of Figure~\ref{fig:overview}, the selected compact star clusters span about $1.5-2\,{\rm mags}$ in color space, which could be indicative of variations in age and metal properties. In the following, we try to characterize these color variations in terms of variations in these physical parameters.

For this, we fit the same SSP models as described in Section~\ref{sec:sedfitting} to each individual compact star cluster.
We note that while the uncertainties of these measurements for individual compact star clusters are rather large due to their faintness, we still are able to investigate statistically trends of age and metallicity in our sample.
Figure~\ref{fig:variations} shows the individual star clusters on the [F090W]$-$[F150W] vs. [F200W]$-$[F277W] color$-$color diagram color-coded by the best-fit age (blue scale) and metallicity (red scale). Note that each symbol has two colors assigned (metallicity as edge-color and age as face-color). Light symbols correspond to star clusters with lower ages and metallicities, while darker symbols correspond to star clusters with older ages and higher metallicities.
The background shows a Gaussian kernel density estimation (including the uncertainties of the data points) to visualize the most probably density of points. The lines indicate colors derived from the SSP models for different metallicities and ages (running from left to right, dots on lines indicated ages of 0.5, 1, 4, and $9\,{\rm Gyrs}$. Older models result in redder [F090W]$-$[F150W] colors, while more metal rich models result in a reddening of both colors.

The range in colors of the individual compact star clusters are coupled with variations in ages and metallicity. This suggests different formation times and possibly different formation processes (causing different metal contents) of the star clusters in the SMACS0723 environment.
We also note that the observed color (hence metallicity) distribution may be expected to be affected by the ``blue tilt'', caused by a lack of the most massive, metal poor star clusters \citep[see][]{FORBES10,USHER18}.

\begin{figure}[t]
\centering
\includegraphics[angle=0,width=1.0\columnwidth]{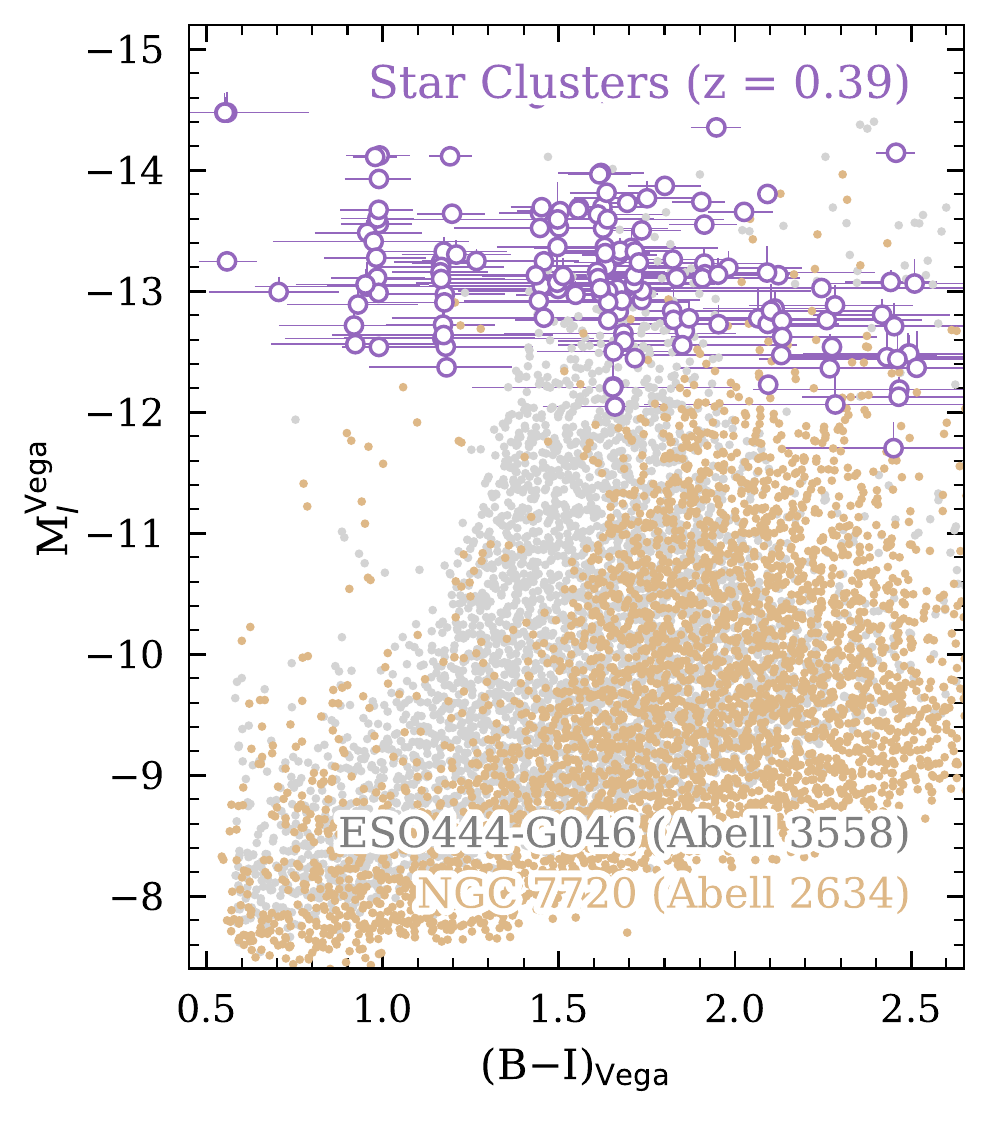}
\caption{
Color vs. magnitude diagram in $(B-I)$ vs. $I$-band absolute luminosity (in Vega system). The brown and gray points correspond to the globulars around BCGs {\it NGC 7720} and {\it ESO444-G046} in the Abell 2634 and 3558 galaxy cluster, respectively, from \citet[][]{HARRIS14}. The latter galaxy cluster is $1.7$ times more massive than the former but still $1.7$ times less massive than SMACS0735. The compact star clusters studied in this work at $z=0.39$ are shown as purple circles. They occupy the bright end at relatively blue $B-I$ colors consistent with the more massive of the two local galaxy cluster shown.
\label{fig:colorlocal}}
\end{figure}

\subsection{Comparison to Local Globular Clusters} \label{sec:localgcs}
 
In the following, we compare the BCG and the identified compact star clusters to systems in the local universe.

The BCG of the SMACS0723 cluster has a stellar mass of $\sim2\times 10^{11}\,{\rm M_\odot}$ and a $V$-band absolute magnitude of $\rm M_V^{Vega} \sim -23.2$\footnote{The $V$-band magnitude was derived from HST photometry and subsequently converted to the Landolt $V$ in Vega system to be consistent with the literature.}, which is at the faint end of the brightness distribution of the BCGs studied in \citet[][]{HARRIS14} ($-24.1 \lesssim \rm M_V^{Vega} \lesssim -23.0$). 
The $V$-band luminosities of our compact star clusters range between $2.5\times 10^6\,{\rm L_\odot}$ and $3.1\times 10^7\,{\rm L_\odot}$, which is about $1-2$ orders of magnitudes brighter than the common turnover luminosity of the globular cluster luminosity function at $\sim 10^5\,{\rm L_\odot}$ \citep[][]{LARSEN01,HARRIS14}. (Note that the lower limit is due to the sensitivity limit of the JWST observations.)
In this luminosity range, the luminosity functions of globulars around local BCGs with similar luminosity find about $100$ globular clusters \citep[][]{HARRIS14}, which is $60\%$ less than what we find in the SMACS0723 cluster field. Several reasons could contribute to this differences.
First, the SMACS0723 galaxy cluster is more massive ($8\times 10^{14}\,{\rm M_\odot}$ as measured by Planck) compared to the Abell clusters ($1-5\times 10^{14}\,{\rm M_\odot}$), which affects the number of globulars. For example, as shown in \citet[][]{HARRIS14}, the luminosity function of globulars around the BCG in Abell 3558 ($4.7\times 10^{14}\,{\rm M_\odot}$) peaks at two times higher number densities as compared to the one of the less massive cluster Abell 1736 ($2.8\times 10^{14}\,{\rm M_\odot}$).
Second, our selection includes the full ICM (over several $100\,{\rm kpc}$, see Figure~\ref{fig:overview}), which may lead to the inclusion of globular clusters around other galaxies in the vicinity of the BCG. Finally, as pointed out in Section~\ref{sec:sizes}, we might also include the tidally stripped compact cores of dwarf galaxies in our sample. Given the above points, the number of compact star clusters is close to what is expected based on local observations.

Note that $>50\%$ of our compact star clusters have $\log(L/L_\odot) > 6.7$ and thus would be considered as superluminous globulars according to \citet[][]{HARRIS14}. Such superluminous objects could be related to ultra compact dwarf galaxies (or the stripped cores thereof) or bridge the gap between globulars and dwarf galaxies. These object have been found in several rich local galaxy clusters \citep[e.g.,][]{MISGELD11,MIESKE12}. SMACS0723 is a massive galaxy cluster at these redshifts (approximately a factor of two more massive than local galaxy clusters studied in Harris et al.), hence the large occurrence of such superluminous globulars would not surprising.

Finally, we compared the color distribution of our compact star clusters to the globulars around local BCGs from \citet[][]{HARRIS14}. Figure~\ref{fig:colorlocal} shows the $(B-I)_{\rm Vega}$ vs. M$_{I}^{\rm Vega}$ color-magnitude diagram for our extracted compact star clusters at $z=0.39$ (purple circles)\footnote{These quantities have been measured on the best-fit de-reddened rest-frame SSP models to the individual star clusters in the Landolt filter system (in Vega magnitudes) to be consistent with other works. Note that the JWST data does not cover the rest-frame $B$ and $I-$bands. Their fluxes are extrapolated from the best-fit models.} as well as globulars around the local galaxies {\it NGC 7720} (gray) and {\it ESO444-G046} (brown), which are the BCGs of the clusters Abell 2634 and 3558, respectively. The former BCG is brighter ($-23.8$ vs. $-23.4\,{\rm Vega\,mag}$) and resides in a more massive cluster ($4.7\times10^{14}$ vs. $1.5\times10^{14}\,{\rm M_\odot}$), which, however, is still $1.7$ times less massive than SMACS0723. 
The compact star clusters studied in this work at $z=0.39$ reside at the bright end of the distribution of local globulars. Their colors are consistent with the colors of globulars in Abell 3558, the more massive one of the two local galaxy clusters shown in Figure~\ref{fig:colorlocal}. This is consistent with the idea that the brightness of the globular clusters scales with the mass of the galaxy clusters.

\subsection{Comparison to Higher Redshift Globulars}\label{sec:highz}
Recent studies by \citet[][]{MOWLA22} and \citet[][]{CLAEYSSENS22} examined the properties of individual globular star clusters in SMACS0723 field of lensed galaxies at higher redshifts.
The significant magnification (up to $100\times$) due to lensing allows the authors of these studies to probe fainter star clusters in the mass range of $10^5-10^7\,{\rm M_\odot}$. However, the wavelength range probed by the JWST NIRCam filters is bluer due to the higher redshift of the sources, which may affect how well different properties of the clusters can be constrained with the given photometry.
The \citet[][]{CLAEYSSENS22} study finds relatively young ages of 100s of Myrs up to $\sim1\,{\rm Gyr}$ for the studied star clusters around $18$ galaxies in the redshift range $1.3 < z < 7.7$. The star clusters are consistent local globulars in terms of their size and stellar mass. These ages are $\sim 1-2\,{\rm Gyr}$ younger than what is measured for our BCG star clusters at $3\sigma$, which is in line with the formation of star clusters over cosmic time. 
On the other hand, \citet[][]{MOWLA22} study in detail a system dubbed as the ``Sparkler'' at $z=1.378$ and find ages of $3-5\,{\rm Gyrs}$ for $6$ out of the $12$ star clusters. This would indicate that these star clusters have been formed during the early reionization phase of the intergalactic medium at $z=8-11$, which would be an interesting but also extreme scenario for cluster formation as understood by current models and other observations of lensed galaxies at high redshifts (see discussion in \citealt[][]{CLAEYSSENS22} and references therein as well as, for example, \citealt[][]{VANZELLA17}). We note that Claeyssens et al. find much younger ages ($\sim 100\,{\rm Myrs}$) for the same star clusters. Understanding this disagreement is beyond the scope of our paper, however, there could be multiple reasons such as aperture vs. PSF fitting photometer methods, the inclusion of galactic extinction, differences in SSP models and SED fitting codes, as well as differences in the JWST photometric calibration references.

\subsection{A Final Word on Milky Way Stars}\label{sec:stars}
Due to their point-like nature and faintness, cool dwarf stars in our Milky Way could be mistaken as globular clusters in the $z=0.39$ host galaxies. A significant contamination of our sample by such stars is, however, ruled out by two reasons.

First, as shown in Figure~\ref{fig:fitresults}, black body models (representing cool Milky Way dwarf stars from $3000-3400\,{\rm K}$) are not consistent with the extracted photometry of our sources. Specifically, the mid-infrared peak is redshifted consistent with the redshift of the cluster galaxies ($z=0.39$). A combination of black body models ({\it i.e.} stars of different temperatures) would be necessary to explain the observed photometry. We note that due to the stacking analysis, stars of different temperatures can be combined, leading to a similar SED as the stacked photometry of the star clusters in our sample. As an additional test, we therefore compared the photometry of randomly picked star clusters to point sources ({\it i.e.} stars) at comparable faint magnitudes selected from the parallel NIRCam module off the center of SMACS0723. We find that the mid-infrared peak emission between the two samples is significantly offset (as a results of the redshift), suggesting negligible contribution of stars to the stack.

Second, the predicted number of stars in the NIRCam FoV is low compared to the selected compact star clusters. Specifically, we predicted the number of stars and brown dwarfs from the \citet[][]{WAINSCOAT92} as well as the \citet[][]{KIRKPATRICK21} models. The former model is in good agreement with the number of stars in the COSMOS field \citep[][]{SCOVILLE07} as shown in \citet[][]{FAJARDOACOSTA22}, however, it does not include consistently M, L, and T spectral types, which start to contribute significantly at fainter magnitudes. The expected number of stars from the Wainscoat and Kirkpatrick models are shown on the left panel of Figure~\ref{fig:properties} as blue histogram and dark blue horizontal lines, respectively. We would expect on the order of $\sim20$ stars and dwarfs over the full FoV of NIRCam, and approximately $5$ over the area where the compact star clusters are found (see Figure~\ref{fig:overview}).

The negligible contribution of Milky Way stars to our sample is also suggested by comparing the number of selected star clusters in the FoVs of the two NIRCam modules. Specifically, the parallel module off-center of the SMACS0723 cluster should be dominated by Milky Way stars. Repeating the sample selection on that field results in only $\sim10$ point sources at similar magnitudes as the selected star clusters. Conservatively assuming that these are all stars and that the stellar density is the same in both FoVs, we would expect $<10$ out of the $178$ selected compact star clusters to be Milky Way stars.

Vice versa, we note that globulars and dwarf galaxies in and around our Milky Way would be easily resolved by JWST and can therefore be ruled out.

\section{Summary and Conclusions}\label{sec:end}

JWST's depth and resolution provides the exciting opportunity to study the faintest and most compact star clusters in our universe. Here, we use the NIRCam early release observations of the SMACS0723 galaxy cluster at $z=0.39$ to study a population of $178$ compact star clusters residing in the intracluster region.

We extract the photometry of these point sources with the \textsc{Tractor} software using robust PSF measurements. We note that there is considerable Galactic dust extinction towards the SMACS0723 field, which needs to be corrected for. We find that the compact star clusters occupy a similar location in [F090W]$-$[F150W] vs. [F200W]$-$[F277W] color space as quiescent galaxies at the same redshifts (Figure~\ref{fig:properties}). However, the star clusters are more than $2-4$ magnitudes fainter than the detection limits of current wide-field surveys and have sizes that are smaller than $50\,{\rm pc}$ (Figures~\ref{fig:psfs} and~\ref{fig:sizes}).

The photometry of the star clusters is fit with SSP models to obtain constraints on their masses, age, and stellar metallicity (Figure~\ref{fig:fitresults}). Different stellar models are tested and the results are found to be robust under the different choices. We derive metallicities between $0.2-0.3\,{\rm Z_{\odot}}$ and robustly exclude $<0.2\,{\rm Z_{\odot}}$ and solar/super-solar metallicities at $4\sigma$ level. For the ages, we find $1.5^{+0.5}_{-0.5}\,{\rm Gyrs}$, ruling out ages $>5\,{\rm Gyrs}$ at $4\sigma$ confidence. The photometry is, however, consistent with older ages up to $9\,{\rm Gyrs}$ at $5\sigma$ level, mainly due to the relatively flat relation between color and age at ages $>3\,{\rm Gyrs}$.
In addition, we find indications of variations of age and metallicity in this cluster population (Figure~\ref{fig:variations}).

Assuming the M/L ratios of the best fit SSP models, we derive stellar masses for the star clusters of $2.4^{+3.0}_{-1.5}\times 10^6\,{\rm M_{\odot}}$ (Figure~\ref{fig:properties}). We note that this M/L ratio is similar to the one of quiescent galaxies at the same redshift and consistent with models and observed ratios of local globulars. The stellar masses of our compact star clusters lie at the high-mass end of average masses of local globulars and there could be overlap with masses of local dwarf galaxies (Figure~\ref{fig:sizes}).

All in all, this suggests that these objects are likely young to middle-aged, compact star clusters (sizes smaller than $50\,{\rm pc}$) with formation times at $z=0.5-0.7$. Such a young age and modest metallicity support a scenario where the star clusters have been formed recently and then were stripped away several $100\,{\rm kpc}$ (projected) from their host galaxies due to the interactions in the cluster field. On the other hand, these clusters could have been formed recently in cold flows onto the cluster core \citep[e.g.,][]{HOLTZMAN92}. Related to this, the recent work by \citet[][]{LEE22} studies the spatial distribution of these star clusters and indicates that they follow closely the intercluster light and trace well the dark matter structure of the galaxy cluster. This would be more in support of a scenario in which the star clusters are stripped from their host galaxies in this highly interactive environment.
A narrower sampling of the $1-5\,{\rm \mu m}$ wavelength with photometry or spectroscopy could help in determining more robust ages to place more stringent constraints on the formation scenario.

The star clusters probed here are $1-2$ orders of magnitudes brighter than the turnover of the globular cluster luminosity function measured around BCGs in local galaxy clusters. The lower end of the luminosity distribution is limited by JWST's sensitivity. At these luminosities, the clusters are in the regime of superluminous globulars as defined in \citet[][]{HARRIS14}, however, it is to note that SMACS0723 is relatively massive compared to local galaxy clusters (although its BCG is at the lower end of the luminosity distribution of local BCG probed in Harris et al.). Interestingly, the $(B-I)$ color distribution of compact star clusters in SMACS0723 is similar to the one of globulars in the $\sim1.7\times$ less massive Abell 3558 cluster, one of the most massive clusters studied in that work (Figure~\ref{fig:colorlocal}).

We note that the star cluster here are, at {\it maximal} sizes of $50\,{\rm pc}$ and masses at the high end of local globulars, among the most compact stellar agglomerations (Figure~\ref{fig:sizes}). Compact star clusters or globulars are a likely explanation of these objects, however, we note that the stripped compact cores of dwarf galaxies could be an alternative explanation (see for example \citet[][]{IDETA04} for a discussion of the origin of $\omega$ Centauri). For example, number estimates based on the Illustris simulation \citep[][]{GENEL14,VOGELSBERGER14a,VOGELSBERGER14b} suggest several hundreds of dwarf galaxies at masses of $10^7\,{\rm M_{\odot}}$ in clusters with $M_{\rm 200} > 5\times 10^{13}\,{\rm M_{\odot}}$, including SMACS0723 at a mass of $8.4\times10^{14}\,{\rm M_{\odot}}$ \citep[][]{MISTANI16,COE19}. Simulations suggest that a dwarf galaxy could lose $\sim90\%$ of its mass during the first few pericenter passages around its host galaxy \citep[e.g.,][]{IDETA04}, making their masses consistent with the ones measured for the clusters (Figure~\ref{fig:sizes}). If a fraction of the expected number of dwarf galaxies gets stripped in the cluster environment, compact cores of these dwarfs could become comparable in number with the compact star clusters seen here.


\begin{acknowledgments}
{\it Acknowledgments:}
We thank Sergio Fajardo-Acosta, Alastair Edge, and Jessica Krick for fruitful discussions on the topic of stars and globular clusters. We also thank the anonymous referee for the helpful input, which significantly improved the quality of this manuscript.
This research is partially funded by the Joint Survey Processing effort at IPAC/Caltech through NASA grant NNN12AA01C.
This research made use of the NASA/IPAC Extragalactic Database (NED), which is operated by the Jet Propulsion Laboratory, California Institute of Technology, under contract with the National Aeronautics and Space Administration.
The Cosmic Dawn Center (DAWN) is funded by the Danish National Research Foundation under grant No. 140.
This work has been based on observations made with the NASA/ESA/CSA James Webb Space Telescope. The data were obtained from the Mikulski Archive for Space Telescopes at the Space Telescope Science Institute, which is operated by the Association of Universities for Research in Astronomy, Inc., under NASA contract NAS 5-03127 for JWST. These observations are associated with JWST programs 2736. The authors acknowledge the ERO teams led by Klaus M. Pontoppidan for developing their observing programs with a zero-exclusive-access period. 
\end{acknowledgments}

%

\vspace{5mm}
\facilities{JWST (NIRCam), VLT (MUSE)}


\software{ \texttt{astropy} \citep{ASTROPY13,ASTROPY18}, 
\texttt{EAZY} \citep[][]{BRAMMER08}, 
\texttt{FSPS} \citep[][]{CONROY09,CONROY10},
\texttt{GALAXEV} \citep[][]{BRUZUALCHARLOT03},
\texttt{Grizli} \citep[][]{BRAMMER21},
\texttt{Source Extractor} \citep{Bertin1996}
          }




\newpage
\begin{center}
\begin{longtable*}{l l l l l l l l l l}
\caption{Coordinates, NIRCam fluxes, and stellar masses for all extracted compact star clusters. All fluxes are corrected for Galactic extinction.  The errors in masses are derived from the photometry only and do not include uncertainties in the M/L ratio.} \label{tab:photo} \\
\hline\hline
\colhead{ID} & \colhead{R.A.} & \colhead{Decl.} & \colhead{F090W} & \colhead{F150W} & \colhead{F200W} & \colhead{F277W} & \colhead{F356W} & \colhead{F444W} & \colhead{log(M/M$_{\odot}$)} \\[-0.2cm]
\colhead{} & \colhead{(J2000)} & \colhead{(J2000)} & \colhead{(nJy)} & \colhead{(nJy)} & \colhead{(nJy)} & \colhead{(nJy)} & \colhead{(nJy)} & \colhead{(nJy)} & \colhead{}\\
\hline 
\endfirsthead
\multicolumn{10}{c}{\tablename\ \thetable{} -- continued} \\
\hline\hline
\colhead{ID} & \colhead{R.A.} & \colhead{Decl.} & \colhead{F090W} & \colhead{F150W} & \colhead{F200W} & \colhead{F277W} & \colhead{F356W} & \colhead{F444W} & \colhead{log(M/M$_{\odot}$)} \\[-0.2cm]
\colhead{} & \colhead{(J2000)} & \colhead{(J2000)} & \colhead{(nJy)} & \colhead{(nJy)} & \colhead{(nJy)} & \colhead{(nJy)} & \colhead{(nJy)} & \colhead{(nJy)} & \colhead{}\\
\hline 
\endhead
\hline \multicolumn{10}{r}{continued on next page}\\
\endfoot
\hline \hline
\endlastfoot
SC\_000 & 110.83478 & -73.45518 & $9.4 \pm 1.2$ & $22.7 \pm 0.3$ & $20.3 \pm 0.3$ & $21.6 \pm 1.2$ & $12.0 \pm 1.1$ & $11.2 \pm 1.5$ & $6.73 \pm 0.05$\\ 
SC\_001 & 110.83487 & -73.45463 & $20.5 \pm 1.2$ & $29.2 \pm 0.5$ & $33.8 \pm 0.7$ & $30.7 \pm 0.6$ & $15.4 \pm 1.1$ & $13.4 \pm 1.5$ & $7.02 \pm 0.02$\\ 
SC\_002 & 110.83238 & -73.45431 & $34.9 \pm 1.2$ & $57.5 \pm 0.8$ & $70.3 \pm 0.7$ & $67.7 \pm 0.5$ & $39.6 \pm 0.5$ & $33.7 \pm 0.5$ & $7.25 \pm 0.02$\\ 
SC\_003 & 110.83262 & -73.45475 & $29.8 \pm 1.2$ & $29.8 \pm 0.7$ & $33.3 \pm 0.7$ & $31.5 \pm 1.1$ & $17.3 \pm 0.7$ & $12.8 \pm 1.5$ & $6.52 \pm 0.02$\\ 
SC\_004 & 110.83263 & -73.45367 & $12.8 \pm 1.2$ & $20.8 \pm 0.7$ & $14.7 \pm 0.5$ & $10.6 \pm 1.2$ & $11.1 \pm 0.7$ & $4.5 \pm 1.5$ & $6.68 \pm 0.04$\\ 
SC\_005 & 110.82962 & -73.45353 & $16.4 \pm 1.2$ & $16.7 \pm 0.5$ & $19.2 \pm 0.5$ & $12.2 \pm 0.8$ & $9.4 \pm 0.6$ & $7.6 \pm 1.5$ & $6.59 \pm 0.03$\\ 
SC\_006 & 110.83044 & -73.45322 & $15.5 \pm 1.2$ & $20.0 \pm 0.5$ & $22.5 \pm 0.7$ & $16.3 \pm 1.2$ & $11.1 \pm 0.7$ & $6.4 \pm 1.5$ & $6.94 \pm 0.04$\\ 
SC\_007 & 110.82800 & -73.45331 & $3.2 \pm 1.2$ & $12.2 \pm 0.7$ & $14.0 \pm 0.7$ & $18.9 \pm 0.5$ & $16.6 \pm 1.1$ & $15.7 \pm 1.5$ & $5.75 \pm 0.14$\\ 
SC\_008 & 110.82647 & -73.45389 & $23.8 \pm 1.2$ & $20.6 \pm 0.5$ & $19.7 \pm 0.6$ & $25.2 \pm 0.8$ & $9.5 \pm 1.1$ & $8.6 \pm 1.5$ & $6.27 \pm 0.02$\\ 
SC\_009 & 110.82579 & -73.45360 & $21.7 \pm 1.2$ & $26.3 \pm 0.5$ & $27.0 \pm 0.4$ & $24.6 \pm 0.7$ & $15.0 \pm 1.1$ & $11.7 \pm 0.7$ & $6.42 \pm 0.02$\\ 
SC\_010 & 110.82474 & -73.45357 & $5.3 \pm 1.2$ & $14.9 \pm 0.4$ & $18.9 \pm 0.4$ & $13.0 \pm 1.2$ & $11.4 \pm 1.1$ & $7.8 \pm 1.5$ & $6.62 \pm 0.09$\\ 
SC\_011 & 110.82139 & -73.45397 & $17.0 \pm 1.2$ & $18.6 \pm 0.6$ & $18.0 \pm 0.3$ & $14.9 \pm 0.5$ & $9.6 \pm 0.7$ & $5.9 \pm 1.5$ & $6.46 \pm 0.03$\\ 
SC\_012 & 110.82147 & -73.45443 & $6.3 \pm 1.2$ & $16.8 \pm 0.8$ & $14.9 \pm 0.7$ & $12.8 \pm 1.2$ & $11.5 \pm 1.1$ & $1.8 \pm 1.5$ & $6.65 \pm 0.10$\\ 
SC\_013 & 110.82276 & -73.45429 & $7.5 \pm 1.2$ & $16.6 \pm 0.7$ & $20.8 \pm 0.4$ & $15.2 \pm 1.2$ & $2.0 \pm 1.1$ & $5.4 \pm 1.5$ & $6.55 \pm 0.07$\\ 
SC\_014 & 110.82239 & -73.45520 & $14.7 \pm 1.2$ & $17.8 \pm 0.8$ & $22.2 \pm 0.6$ & $20.7 \pm 1.2$ & $12.0 \pm 1.1$ & $13.3 \pm 1.5$ & $6.50 \pm 0.04$\\ 
SC\_015 & 110.82412 & -73.45540 & $26.1 \pm 1.2$ & $42.1 \pm 0.2$ & $42.7 \pm 0.3$ & $47.2 \pm 1.2$ & $27.2 \pm 1.0$ & $21.7 \pm 1.5$ & $6.36 \pm 0.02$\\ 
SC\_016 & 110.82460 & -73.45579 & $15.2 \pm 1.2$ & $16.6 \pm 0.5$ & $18.5 \pm 0.5$ & $19.5 \pm 1.1$ & $7.3 \pm 1.1$ & $7.4 \pm 1.1$ & $6.13 \pm 0.03$\\ 
SC\_017 & 110.82441 & -73.45585 & $8.1 \pm 1.2$ & $14.6 \pm 0.6$ & $13.0 \pm 0.5$ & $16.9 \pm 0.5$ & $15.3 \pm 0.9$ & $13.5 \pm 1.5$ & $5.92 \pm 0.06$\\ 
SC\_018 & 110.82575 & -73.45549 & $20.8 \pm 1.2$ & $28.0 \pm 0.6$ & $21.5 \pm 0.5$ & $21.9 \pm 1.2$ & $16.8 \pm 1.1$ & $16.7 \pm 1.5$ & $6.16 \pm 0.02$\\ 
SC\_019 & 110.82953 & -73.45540 & $18.3 \pm 1.2$ & $21.3 \pm 0.8$ & $28.3 \pm 0.2$ & $33.8 \pm 1.2$ & $20.3 \pm 1.1$ & $24.1 \pm 1.5$ & $6.52 \pm 0.02$\\ 
SC\_020 & 110.83033 & -73.45513 & $26.0 \pm 1.2$ & $34.4 \pm 0.4$ & $38.6 \pm 0.4$ & $42.2 \pm 1.2$ & $26.4 \pm 1.1$ & $13.8 \pm 1.5$ & $6.43 \pm 0.02$\\ 
SC\_021 & 110.83040 & -73.45523 & $12.0 \pm 1.2$ & $21.5 \pm 0.7$ & $27.5 \pm 0.6$ & $39.3 \pm 1.2$ & $29.3 \pm 1.1$ & $30.0 \pm 1.5$ & $6.34 \pm 0.04$\\ 
SC\_022 & 110.83355 & -73.45528 & $11.5 \pm 1.2$ & $12.9 \pm 0.5$ & $13.0 \pm 0.6$ & $12.9 \pm 1.0$ & $5.3 \pm 1.1$ & $6.3 \pm 1.5$ & $5.95 \pm 0.05$\\ 
SC\_023 & 110.83300 & -73.45532 & $6.8 \pm 1.2$ & $15.5 \pm 0.5$ & $14.4 \pm 0.5$ & $13.7 \pm 1.0$ & $7.1 \pm 0.7$ & $6.2 \pm 1.5$ & $6.63 \pm 0.08$\\ 
SC\_024 & 110.83395 & -73.45505 & $15.7 \pm 0.1$ & $15.8 \pm 0.6$ & $11.1 \pm 0.6$ & $10.1 \pm 1.2$ & $3.8 \pm 1.1$ & $1.2 \pm 1.5$ & $6.22 \pm 0.00$\\ 
SC\_025 & 110.83219 & -73.45462 & $19.3 \pm 1.2$ & $22.8 \pm 0.6$ & $24.3 \pm 0.7$ & $20.5 \pm 0.5$ & $17.1 \pm 1.0$ & $12.7 \pm 1.5$ & $6.37 \pm 0.03$\\ 
SC\_026 & 110.83015 & -73.45389 & $19.9 \pm 0.0$ & $30.4 \pm 0.8$ & $34.9 \pm 0.5$ & $36.0 \pm 0.9$ & $18.9 \pm 0.7$ & $18.3 \pm 1.5$ & $6.18 \pm 0.00$\\ 
SC\_027 & 110.82938 & -73.45402 & $12.7 \pm 1.2$ & $17.0 \pm 0.8$ & $21.8 \pm 0.5$ & $17.4 \pm 1.2$ & $12.8 \pm 1.1$ & $9.3 \pm 1.5$ & $6.49 \pm 0.04$\\ 
SC\_028 & 110.83031 & -73.45406 & $18.4 \pm 1.2$ & $14.4 \pm 0.7$ & $14.7 \pm 0.6$ & $19.7 \pm 1.2$ & $9.2 \pm 1.1$ & $10.7 \pm 1.5$ & $6.11 \pm 0.03$\\ 
SC\_029 & 110.83097 & -73.45398 & $4.4 \pm 1.2$ & $13.7 \pm 0.8$ & $12.0 \pm 0.5$ & $9.8 \pm 0.4$ & $2.3 \pm 1.1$ & $2.5 \pm 1.5$ & $6.17 \pm 0.12$\\ 
SC\_030 & 110.83071 & -73.45574 & $46.2 \pm 1.2$ & $55.4 \pm 0.4$ & $59.1 \pm 0.7$ & $52.6 \pm 1.2$ & $29.8 \pm 0.5$ & $20.5 \pm 1.5$ & $7.08 \pm 0.01$\\ 
SC\_031 & 110.83011 & -73.45603 & $10.3 \pm 1.2$ & $21.3 \pm 0.8$ & $18.8 \pm 0.6$ & $15.2 \pm 0.3$ & $9.5 \pm 1.1$ & $4.6 \pm 1.5$ & $6.76 \pm 0.04$\\ 
SC\_032 & 110.83222 & -73.45585 & $10.8 \pm 1.2$ & $13.2 \pm 0.4$ & $14.6 \pm 0.5$ & $13.4 \pm 1.2$ & $7.1 \pm 0.1$ & $2.9 \pm 1.5$ & $6.76 \pm 0.04$\\ 
SC\_033 & 110.82036 & -73.45593 & $5.2 \pm 1.2$ & $12.7 \pm 0.8$ & $13.1 \pm 0.5$ & $15.8 \pm 0.6$ & $7.3 \pm 1.0$ & $6.0 \pm 1.5$ & $6.53 \pm 0.09$\\ 
SC\_034 & 110.82258 & -73.45596 & $10.0 \pm 1.2$ & $12.7 \pm 0.3$ & $12.6 \pm 0.4$ & $8.3 \pm 1.2$ & $3.7 \pm 1.1$ & $3.3 \pm 1.5$ & $6.68 \pm 0.05$\\ 
SC\_035 & 110.80869 & -73.45551 & $15.6 \pm 1.2$ & $18.8 \pm 0.5$ & $20.8 \pm 0.5$ & $19.6 \pm 0.6$ & $10.4 \pm 1.1$ & $6.6 \pm 1.5$ & $6.37 \pm 0.04$\\ 
SC\_036 & 110.81668 & -73.45396 & $15.0 \pm 1.2$ & $10.1 \pm 0.6$ & $13.3 \pm 0.5$ & $16.7 \pm 1.2$ & $7.7 \pm 1.1$ & $8.0 \pm 1.5$ & $6.12 \pm 0.04$\\ 
SC\_037 & 110.81672 & -73.45406 & $3.9 \pm 1.2$ & $21.2 \pm 0.8$ & $19.9 \pm 0.4$ & $16.8 \pm 0.9$ & $7.1 \pm 0.6$ & $3.5 \pm 1.5$ & $6.45 \pm 0.14$\\ 
SC\_038 & 110.81860 & -73.45359 & $6.8 \pm 1.2$ & $20.8 \pm 0.7$ & $23.1 \pm 0.6$ & $24.1 \pm 1.2$ & $14.1 \pm 1.1$ & $7.6 \pm 1.5$ & $6.89 \pm 0.07$\\ 
SC\_039 & 110.82047 & -73.45328 & $31.2 \pm 1.2$ & $32.3 \pm 0.4$ & $32.5 \pm 0.4$ & $31.2 \pm 0.5$ & $20.6 \pm 0.5$ & $8.4 \pm 1.5$ & $6.38 \pm 0.01$\\ 
SC\_040 & 110.82181 & -73.45330 & $12.8 \pm 1.2$ & $19.8 \pm 0.1$ & $15.7 \pm 0.2$ & $16.9 \pm 1.2$ & $15.1 \pm 1.1$ & $15.6 \pm 1.5$ & $6.12 \pm 0.04$\\ 
SC\_041 & 110.82086 & -73.45378 & $2.8 \pm 1.2$ & $11.5 \pm 0.7$ & $11.6 \pm 0.4$ & $8.6 \pm 1.2$ & $7.5 \pm 0.7$ & $2.9 \pm 1.5$ & $6.32 \pm 0.18$\\ 
SC\_042 & 110.82852 & -73.45620 & $12.0 \pm 0.1$ & $17.6 \pm 0.4$ & $17.1 \pm 0.6$ & $16.0 \pm 0.4$ & $8.8 \pm 1.1$ & $5.9 \pm 1.5$ & $6.37 \pm 0.00$\\ 
SC\_043 & 110.82804 & -73.45646 & $17.9 \pm 1.2$ & $20.7 \pm 0.8$ & $15.5 \pm 0.3$ & $15.2 \pm 1.2$ & $8.4 \pm 1.1$ & $6.9 \pm 1.5$ & $6.26 \pm 0.03$\\ 
SC\_044 & 110.82936 & -73.45614 & $18.6 \pm 1.2$ & $17.1 \pm 0.5$ & $18.0 \pm 0.5$ & $20.0 \pm 1.2$ & $16.6 \pm 1.1$ & $21.3 \pm 1.5$ & $6.28 \pm 0.03$\\ 
SC\_045 & 110.82436 & -73.45603 & $8.0 \pm 1.2$ & $14.9 \pm 0.7$ & $16.0 \pm 0.6$ & $13.3 \pm 0.8$ & $6.9 \pm 1.1$ & $1.9 \pm 1.5$ & $6.39 \pm 0.06$\\ 
SC\_046 & 110.82245 & -73.45393 & $11.0 \pm 1.2$ & $13.9 \pm 0.3$ & $11.9 \pm 0.5$ & $11.0 \pm 0.3$ & $4.6 \pm 0.7$ & $4.4 \pm 1.5$ & $6.59 \pm 0.04$\\ 
SC\_047 & 110.82245 & -73.45386 & $9.5 \pm 1.2$ & $14.6 \pm 0.7$ & $14.3 \pm 0.6$ & $14.5 \pm 0.6$ & $18.8 \pm 0.4$ & $10.9 \pm 1.5$ & $5.99 \pm 0.06$\\ 
SC\_048 & 110.82226 & -73.45405 & $6.0 \pm 1.2$ & $7.1 \pm 0.6$ & $8.2 \pm 0.5$ & $7.4 \pm 1.2$ & $8.3 \pm 1.1$ & $1.9 \pm 1.5$ & $5.95 \pm 0.09$\\ 
SC\_049 & 110.81272 & -73.45459 & $19.8 \pm 1.2$ & $25.4 \pm 0.8$ & $25.6 \pm 0.6$ & $21.9 \pm 1.2$ & $12.9 \pm 1.1$ & $9.5 \pm 1.5$ & $7.05 \pm 0.02$\\ 
SC\_050 & 110.82056 & -73.45527 & $10.7 \pm 1.2$ & $20.9 \pm 0.8$ & $20.4 \pm 0.6$ & $24.2 \pm 1.2$ & $10.7 \pm 0.7$ & $12.8 \pm 1.5$ & $6.74 \pm 0.04$\\ 
SC\_051 & 110.83121 & -73.45448 & $10.5 \pm 1.2$ & $18.1 \pm 0.8$ & $18.0 \pm 0.6$ & $17.6 \pm 1.2$ & $10.5 \pm 1.1$ & $9.6 \pm 1.5$ & $6.80 \pm 0.05$\\ 
SC\_052 & 110.83858 & -73.45495 & $12.0 \pm 1.2$ & $19.7 \pm 0.8$ & $18.9 \pm 0.4$ & $17.0 \pm 1.2$ & $10.8 \pm 1.1$ & $6.6 \pm 1.5$ & $6.36 \pm 0.05$\\ 
SC\_053 & 110.83849 & -73.45514 & $19.3 \pm 1.2$ & $24.8 \pm 0.6$ & $22.9 \pm 0.5$ & $22.8 \pm 0.7$ & $13.3 \pm 0.7$ & $7.7 \pm 1.5$ & $6.17 \pm 0.02$\\ 
SC\_054 & 110.83820 & -73.45532 & $9.5 \pm 1.2$ & $8.4 \pm 0.6$ & $10.7 \pm 0.7$ & $11.2 \pm 1.2$ & $5.0 \pm 1.1$ & $4.8 \pm 1.5$ & $5.93 \pm 0.06$\\ 
SC\_055 & 110.83400 & -73.45490 & $6.1 \pm 1.2$ & $9.3 \pm 0.8$ & $15.3 \pm 0.6$ & $15.0 \pm 1.1$ & $6.4 \pm 1.1$ & $7.5 \pm 1.5$ & $6.75 \pm 0.10$\\ 
SC\_056 & 110.83486 & -73.45413 & $7.2 \pm 1.2$ & $11.4 \pm 0.6$ & $12.0 \pm 0.3$ & $9.5 \pm 1.2$ & $7.8 \pm 1.1$ & $8.4 \pm 1.5$ & $6.63 \pm 0.07$\\ 
SC\_057 & 110.82368 & -73.45312 & $7.7 \pm 1.2$ & $11.5 \pm 0.5$ & $16.3 \pm 0.2$ & $13.4 \pm 0.6$ & $8.0 \pm 1.1$ & $2.2 \pm 1.5$ & $6.78 \pm 0.07$\\ 
SC\_058 & 110.82875 & -73.45347 & $5.8 \pm 1.2$ & $8.2 \pm 0.6$ & $10.7 \pm 0.5$ & $8.6 \pm 1.2$ & $5.3 \pm 1.1$ & $4.1 \pm 1.5$ & $6.20 \pm 0.07$\\ 
SC\_059 & 110.84541 & -73.45423 & $9.5 \pm 1.2$ & $17.9 \pm 0.5$ & $15.5 \pm 0.7$ & $11.1 \pm 1.2$ & $5.6 \pm 1.1$ & $7.5 \pm 0.3$ & $6.72 \pm 0.06$\\ 
SC\_060 & 110.84582 & -73.45393 & $6.3 \pm 1.2$ & $12.8 \pm 0.6$ & $11.3 \pm 0.7$ & $10.0 \pm 1.2$ & $7.9 \pm 1.0$ & $6.0 \pm 0.5$ & $6.58 \pm 0.08$\\ 
SC\_061 & 110.84403 & -73.45389 & $3.9 \pm 1.1$ & $10.4 \pm 0.8$ & $11.2 \pm 0.2$ & $11.9 \pm 1.2$ & $6.3 \pm 1.1$ & $4.7 \pm 1.5$ & $6.68 \pm 0.14$\\ 
SC\_062 & 110.84351 & -73.45305 & $9.3 \pm 1.2$ & $14.2 \pm 0.8$ & $12.2 \pm 0.7$ & $11.7 \pm 1.2$ & $6.9 \pm 1.1$ & $3.0 \pm 1.5$ & $6.72 \pm 0.06$\\ 
SC\_063 & 110.84426 & -73.45268 & $7.1 \pm 0.1$ & $19.4 \pm 0.5$ & $19.6 \pm 0.5$ & $19.7 \pm 1.2$ & $9.7 \pm 0.8$ & $6.6 \pm 1.5$ & $6.56 \pm 0.01$\\ 
SC\_064 & 110.84474 & -73.45277 & $10.8 \pm 1.2$ & $18.6 \pm 0.8$ & $19.3 \pm 0.6$ & $21.3 \pm 1.2$ & $12.0 \pm 1.1$ & $8.3 \pm 1.5$ & $6.83 \pm 0.05$\\ 
SC\_065 & 110.84351 & -73.45347 & $2.0 \pm 1.2$ & $13.4 \pm 0.7$ & $14.9 \pm 0.7$ & $11.8 \pm 1.2$ & $6.6 \pm 0.0$ & $3.1 \pm 1.5$ & $5.23 \pm 0.22$\\ 
SC\_066 & 110.83753 & -73.45430 & $10.7 \pm 1.2$ & $16.5 \pm 0.5$ & $12.7 \pm 0.7$ & $18.2 \pm 1.2$ & $12.1 \pm 0.5$ & $7.0 \pm 1.5$ & $6.04 \pm 0.05$\\ 
SC\_067 & 110.82613 & -73.45409 & $4.1 \pm 1.2$ & $24.4 \pm 0.7$ & $21.5 \pm 0.6$ & $28.9 \pm 1.2$ & $17.8 \pm 1.1$ & $28.6 \pm 1.5$ & $6.69 \pm 0.13$\\ 
SC\_068 & 110.83132 & -73.45512 & $4.6 \pm 1.2$ & $6.0 \pm 0.8$ & $10.7 \pm 0.6$ & $5.3 \pm 1.1$ & $1.9 \pm 1.1$ & $0.8 \pm 1.5$ & $6.06 \pm 0.11$\\ 
SC\_069 & 110.82314 & -73.45583 & $42.1 \pm 1.2$ & $53.0 \pm 0.0$ & $37.3 \pm 0.1$ & $32.8 \pm 1.1$ & $35.2 \pm 1.1$ & $29.7 \pm 1.5$ & $6.50 \pm 0.02$\\ 
SC\_070 & 110.82385 & -73.45596 & $17.6 \pm 1.2$ & $26.8 \pm 0.8$ & $29.5 \pm 0.7$ & $28.9 \pm 1.2$ & $18.3 \pm 1.0$ & $13.9 \pm 1.5$ & $6.57 \pm 0.03$\\ 
SC\_071 & 110.83472 & -73.45482 & $6.5 \pm 1.2$ & $12.4 \pm 0.7$ & $11.3 \pm 0.4$ & $9.5 \pm 1.2$ & $5.9 \pm 1.1$ & $8.6 \pm 1.5$ & $6.57 \pm 0.08$\\ 
SC\_072 & 110.84418 & -73.45358 & $7.7 \pm 1.2$ & $7.2 \pm 0.2$ & $5.2 \pm 0.7$ & $7.2 \pm 1.2$ & $4.4 \pm 1.1$ & $5.7 \pm 1.5$ & $5.74 \pm 0.05$\\ 
SC\_073 & 110.84534 & -73.45344 & $8.4 \pm 1.2$ & $13.2 \pm 0.7$ & $15.0 \pm 0.7$ & $13.5 \pm 1.0$ & $6.5 \pm 1.1$ & $5.3 \pm 1.5$ & $6.59 \pm 0.06$\\ 
SC\_074 & 110.80991 & -73.45428 & $7.1 \pm 1.2$ & $5.8 \pm 0.8$ & $7.3 \pm 0.6$ & $6.1 \pm 1.2$ & $5.8 \pm 1.1$ & $-0.1 \pm 1.5$ & $6.47 \pm 0.07$\\ 
SC\_075 & 110.81175 & -73.45398 & $4.1 \pm 1.2$ & $11.9 \pm 0.4$ & $13.9 \pm 0.7$ & $10.8 \pm 1.2$ & $7.9 \pm 1.0$ & $3.8 \pm 1.5$ & $6.48 \pm 0.15$\\ 
SC\_076 & 110.81713 & -73.45409 & $12.9 \pm 1.2$ & $22.3 \pm 0.0$ & $17.0 \pm 0.6$ & $17.3 \pm 1.2$ & $12.1 \pm 0.7$ & $6.3 \pm 1.5$ & $5.99 \pm 0.04$\\ 
SC\_077 & 110.82002 & -73.45379 & $20.2 \pm 1.2$ & $21.5 \pm 0.5$ & $20.4 \pm 0.7$ & $-0.7 \pm 1.2$ & $-1.2 \pm 1.1$ & $-2.9 \pm 1.5$ & $6.92 \pm 0.03$\\ 
SC\_078 & 110.83525 & -73.45489 & $4.9 \pm 1.2$ & $8.4 \pm 0.6$ & $12.0 \pm 0.7$ & $12.5 \pm 1.2$ & $6.9 \pm 1.1$ & $4.3 \pm 1.4$ & $6.24 \pm 0.11$\\ 
SC\_079 & 110.83585 & -73.45481 & $6.6 \pm 1.2$ & $11.6 \pm 0.8$ & $11.6 \pm 0.7$ & $11.3 \pm 1.2$ & $6.2 \pm 0.8$ & $1.7 \pm 1.5$ & $6.30 \pm 0.09$\\ 
SC\_080 & 110.84533 & -73.45445 & $7.7 \pm 1.2$ & $14.5 \pm 0.7$ & $14.1 \pm 0.5$ & $14.0 \pm 1.1$ & $7.5 \pm 1.1$ & $3.9 \pm 1.5$ & $6.75 \pm 0.06$\\ 
SC\_081 & 110.84509 & -73.45449 & $4.9 \pm 1.2$ & $10.0 \pm 0.7$ & $9.8 \pm 0.4$ & $7.9 \pm 1.2$ & $4.6 \pm 0.6$ & $-2.5 \pm 1.5$ & $6.28 \pm 0.11$\\ 
SC\_082 & 110.82796 & -73.45581 & $4.7 \pm 1.2$ & $8.8 \pm 0.8$ & $12.6 \pm 0.6$ & $8.9 \pm 1.2$ & $8.7 \pm 1.1$ & $4.8 \pm 1.5$ & $6.55 \pm 0.12$\\ 
SC\_083 & 110.82717 & -73.45708 & $7.1 \pm 1.2$ & $18.8 \pm 0.6$ & $20.3 \pm 0.5$ & $19.4 \pm 0.7$ & $11.0 \pm 1.0$ & $8.0 \pm 1.5$ & $6.63 \pm 0.07$\\ 
SC\_084 & 110.78770 & -73.45499 & $6.6 \pm 1.2$ & $10.7 \pm 0.6$ & $9.7 \pm 0.6$ & $10.5 \pm 1.2$ & $3.8 \pm 1.1$ & $6.5 \pm 0.5$ & $6.15 \pm 0.07$\\ 
SC\_085 & 110.81329 & -73.45452 & $23.5 \pm 1.2$ & $13.6 \pm 0.8$ & $14.3 \pm 0.5$ & $2.1 \pm 1.2$ & $1.8 \pm 1.1$ & $-0.4 \pm 1.5$ & $6.98 \pm 0.02$\\ 
SC\_086 & 110.82160 & -73.45648 & $7.5 \pm 1.2$ & $8.3 \pm 0.8$ & $13.7 \pm 0.7$ & $10.9 \pm 0.8$ & $5.8 \pm 1.0$ & $7.8 \pm 1.5$ & $6.37 \pm 0.07$\\ 
SC\_087 & 110.81876 & -73.45635 & $10.8 \pm 0.1$ & $11.7 \pm 0.8$ & $15.3 \pm 0.7$ & $10.6 \pm 1.2$ & $5.0 \pm 1.0$ & $5.3 \pm 1.5$ & $6.75 \pm 0.00$\\ 
SC\_088 & 110.81799 & -73.45675 & $2.8 \pm 1.2$ & $8.5 \pm 0.8$ & $7.8 \pm 0.5$ & $9.5 \pm 0.4$ & $5.3 \pm 1.1$ & $2.3 \pm 1.5$ & $5.73 \pm 0.19$\\ 
SC\_089 & 110.81732 & -73.45681 & $11.4 \pm 1.2$ & $12.8 \pm 0.7$ & $13.3 \pm 0.4$ & $11.0 \pm 0.7$ & $4.5 \pm 1.1$ & $4.0 \pm 1.5$ & $6.45 \pm 0.04$\\ 
SC\_090 & 110.81038 & -73.45548 & $14.1 \pm 1.2$ & $17.6 \pm 0.7$ & $18.7 \pm 0.7$ & $21.8 \pm 1.2$ & $12.9 \pm 1.1$ & $12.1 \pm 1.5$ & $6.16 \pm 0.04$\\ 
SC\_091 & 110.78922 & -73.45409 & $14.0 \pm 0.7$ & $8.6 \pm 0.6$ & $9.6 \pm 0.4$ & $10.2 \pm 0.9$ & $4.7 \pm 1.1$ & $3.1 \pm 1.5$ & $6.02 \pm 0.02$\\ 
SC\_092 & 110.78970 & -73.45433 & $7.7 \pm 1.2$ & $10.8 \pm 0.5$ & $11.2 \pm 0.4$ & $8.8 \pm 1.2$ & $4.7 \pm 0.7$ & $5.5 \pm 1.5$ & $6.64 \pm 0.07$\\ 
SC\_093 & 110.79125 & -73.45385 & $9.8 \pm 1.2$ & $10.2 \pm 0.5$ & $10.7 \pm 0.5$ & $9.8 \pm 1.2$ & $3.5 \pm 1.1$ & $2.4 \pm 1.5$ & $6.33 \pm 0.05$\\ 
SC\_094 & 110.78340 & -73.45303 & $6.7 \pm 0.1$ & $8.2 \pm 0.5$ & $9.5 \pm 0.4$ & $7.4 \pm 1.2$ & $2.8 \pm 1.1$ & $2.2 \pm 1.5$ & $6.57 \pm 0.01$\\ 
SC\_095 & 110.78423 & -73.45257 & $1.9 \pm 1.2$ & $6.7 \pm 0.4$ & $7.7 \pm 0.4$ & $6.8 \pm 1.2$ & $2.8 \pm 1.1$ & $1.2 \pm 1.5$ & $6.41 \pm 0.24$\\ 
SC\_096 & 110.78713 & -73.45280 & $22.1 \pm 1.2$ & $22.9 \pm 0.6$ & $23.7 \pm 0.7$ & $21.4 \pm 1.2$ & $12.1 \pm 0.8$ & $9.5 \pm 1.5$ & $6.64 \pm 0.03$\\ 
SC\_097 & 110.78789 & -73.45194 & $13.2 \pm 1.2$ & $17.7 \pm 0.4$ & $18.5 \pm 0.5$ & $14.6 \pm 1.2$ & $9.5 \pm 1.1$ & $5.8 \pm 1.5$ & $6.86 \pm 0.03$\\ 
SC\_098 & 110.78635 & -73.45047 & $7.2 \pm 1.2$ & $14.7 \pm 0.6$ & $14.0 \pm 0.4$ & $18.4 \pm 1.2$ & $10.8 \pm 0.6$ & $9.2 \pm 1.5$ & $5.88 \pm 0.08$\\ 
SC\_099 & 110.78172 & -73.45070 & $10.3 \pm 1.2$ & $9.8 \pm 0.8$ & $10.8 \pm 0.6$ & $9.1 \pm 0.6$ & $4.5 \pm 1.1$ & $1.8 \pm 1.5$ & $6.35 \pm 0.04$\\ 
SC\_100 & 110.79601 & -73.44790 & $6.0 \pm 1.2$ & $16.9 \pm 0.7$ & $19.2 \pm 0.5$ & $18.7 \pm 1.2$ & $9.8 \pm 0.6$ & $6.0 \pm 1.5$ & $6.83 \pm 0.08$\\ 
SC\_101 & 110.79508 & -73.44825 & $14.7 \pm 1.2$ & $13.2 \pm 0.7$ & $14.7 \pm 0.7$ & $13.0 \pm 1.2$ & $6.2 \pm 0.8$ & $10.0 \pm 1.5$ & $6.30 \pm 0.04$\\ 
SC\_102 & 110.79584 & -73.44829 & $9.4 \pm 1.2$ & $16.7 \pm 0.7$ & $15.7 \pm 0.4$ & $14.3 \pm 1.2$ & $8.9 \pm 1.1$ & $7.0 \pm 1.5$ & $6.32 \pm 0.06$\\ 
SC\_103 & 110.80174 & -73.44977 & $13.2 \pm 1.2$ & $17.9 \pm 0.6$ & $17.8 \pm 0.5$ & $19.1 \pm 0.7$ & $9.5 \pm 1.1$ & $8.6 \pm 1.5$ & $6.07 \pm 0.05$\\ 
SC\_104 & 110.80198 & -73.44992 & $9.3 \pm 1.2$ & $13.8 \pm 0.4$ & $16.7 \pm 0.1$ & $14.3 \pm 1.1$ & $8.0 \pm 1.0$ & $5.7 \pm 1.5$ & $6.75 \pm 0.06$\\ 
SC\_105 & 110.80116 & -73.44958 & $9.9 \pm 1.2$ & $16.3 \pm 0.7$ & $16.0 \pm 0.5$ & $18.8 \pm 0.6$ & $13.8 \pm 1.1$ & $14.2 \pm 1.5$ & $6.00 \pm 0.04$\\ 
SC\_106 & 110.80534 & -73.44852 & $8.6 \pm 1.2$ & $11.5 \pm 0.7$ & $16.0 \pm 0.5$ & $9.4 \pm 0.5$ & $8.0 \pm 0.5$ & $6.0 \pm 1.5$ & $6.29 \pm 0.05$\\ 
SC\_107 & 110.80524 & -73.44870 & $8.0 \pm 1.2$ & $10.6 \pm 0.3$ & $8.2 \pm 0.7$ & $11.9 \pm 1.2$ & $5.5 \pm 0.9$ & $2.2 \pm 1.5$ & $5.75 \pm 0.06$\\ 
SC\_108 & 110.78716 & -73.45552 & $8.9 \pm 1.2$ & $14.5 \pm 0.5$ & $15.0 \pm 0.5$ & $15.0 \pm 1.2$ & $9.2 \pm 1.1$ & $7.6 \pm 1.5$ & $6.28 \pm 0.06$\\ 
SC\_109 & 110.78741 & -73.45586 & $9.9 \pm 1.2$ & $11.8 \pm 0.7$ & $15.5 \pm 0.6$ & $12.2 \pm 1.2$ & $6.4 \pm 0.5$ & $4.2 \pm 0.4$ & $6.72 \pm 0.06$\\ 
SC\_110 & 110.78030 & -73.45425 & $7.6 \pm 1.2$ & $13.3 \pm 0.5$ & $14.7 \pm 0.6$ & $12.7 \pm 0.4$ & $6.1 \pm 0.7$ & $3.1 \pm 1.5$ & $6.55 \pm 0.07$\\ 
SC\_111 & 110.78126 & -73.45340 & $5.6 \pm 1.2$ & $11.9 \pm 0.4$ & $15.3 \pm 0.7$ & $15.7 \pm 0.4$ & $9.0 \pm 1.1$ & $8.3 \pm 1.5$ & $6.76 \pm 0.08$\\ 
SC\_112 & 110.78087 & -73.45291 & $10.7 \pm 1.2$ & $17.9 \pm 0.5$ & $20.0 \pm 0.5$ & $20.6 \pm 1.2$ & $13.2 \pm 0.9$ & $8.3 \pm 1.5$ & $6.75 \pm 0.04$\\ 
SC\_113 & 110.79191 & -73.45265 & $9.3 \pm 1.2$ & $15.2 \pm 0.7$ & $14.4 \pm 0.4$ & $13.2 \pm 1.2$ & $8.5 \pm 0.2$ & $4.0 \pm 1.5$ & $6.27 \pm 0.06$\\ 
SC\_114 & 110.79176 & -73.45274 & $11.4 \pm 1.2$ & $13.2 \pm 0.8$ & $13.5 \pm 0.6$ & $10.8 \pm 1.2$ & $5.9 \pm 1.1$ & $5.0 \pm 1.5$ & $6.66 \pm 0.04$\\ 
SC\_115 & 110.79931 & -73.45132 & $7.9 \pm 1.2$ & $15.4 \pm 0.7$ & $13.7 \pm 0.4$ & $11.7 \pm 0.8$ & $6.4 \pm 1.1$ & $7.9 \pm 1.5$ & $6.65 \pm 0.07$\\ 
SC\_116 & 110.79995 & -73.45124 & $12.0 \pm 1.2$ & $9.8 \pm 0.8$ & $11.8 \pm 0.6$ & $10.0 \pm 1.2$ & $4.6 \pm 0.3$ & $6.4 \pm 1.5$ & $6.09 \pm 0.04$\\ 
SC\_117 & 110.79854 & -73.45139 & $4.4 \pm 1.2$ & $12.8 \pm 0.5$ & $10.2 \pm 0.5$ & $10.9 \pm 1.2$ & $5.4 \pm 0.6$ & $3.4 \pm 1.5$ & $6.41 \pm 0.12$\\ 
SC\_118 & 110.79830 & -73.45146 & $7.1 \pm 1.2$ & $8.5 \pm 0.6$ & $9.3 \pm 0.4$ & $10.9 \pm 0.3$ & $5.2 \pm 0.6$ & $0.2 \pm 1.5$ & $5.87 \pm 0.07$\\ 
SC\_119 & 110.79798 & -73.45131 & $7.7 \pm 1.2$ & $9.6 \pm 0.5$ & $11.5 \pm 0.5$ & $9.5 \pm 1.2$ & $5.3 \pm 1.1$ & $3.6 \pm 1.5$ & $6.25 \pm 0.07$\\ 
SC\_120 & 110.79658 & -73.45191 & $13.1 \pm 1.2$ & $15.0 \pm 0.7$ & $13.5 \pm 0.6$ & $11.4 \pm 1.2$ & $7.4 \pm 0.7$ & $2.6 \pm 1.5$ & $6.25 \pm 0.04$\\ 
SC\_121 & 110.78631 & -73.45810 & $23.0 \pm 1.2$ & $25.2 \pm 0.8$ & $24.1 \pm 0.7$ & $25.6 \pm 1.2$ & $19.6 \pm 1.1$ & $11.2 \pm 1.5$ & $6.26 \pm 0.02$\\ 
SC\_122 & 110.78559 & -73.45643 & $10.1 \pm 1.2$ & $14.8 \pm 0.7$ & $14.9 \pm 0.3$ & $13.9 \pm 1.2$ & $6.0 \pm 1.1$ & $5.0 \pm 1.5$ & $6.75 \pm 0.05$\\ 
SC\_123 & 110.78135 & -73.45553 & $8.2 \pm 1.2$ & $21.7 \pm 0.7$ & $19.7 \pm 0.7$ & $22.2 \pm 1.2$ & $14.3 \pm 0.9$ & $9.3 \pm 1.5$ & $6.86 \pm 0.07$\\ 
SC\_124 & 110.67039 & -73.48220 & $17.8 \pm 1.2$ & $14.8 \pm 0.7$ & $19.1 \pm 0.7$ & $26.5 \pm 1.2$ & $17.0 \pm 1.1$ & $14.8 \pm 1.5$ & $6.85 \pm 0.03$\\ 
SC\_125 & 110.68257 & -73.47759 & $15.3 \pm 1.2$ & $12.4 \pm 0.6$ & $16.0 \pm 0.6$ & $19.9 \pm 1.2$ & $8.7 \pm 1.1$ & $8.9 \pm 1.5$ & $6.79 \pm 0.03$\\ 
SC\_126 & 110.73865 & -73.47510 & $5.6 \pm 1.2$ & $14.7 \pm 0.6$ & $14.0 \pm 0.1$ & $14.0 \pm 1.2$ & $29.0 \pm 1.1$ & $20.4 \pm 1.5$ & $6.35 \pm 0.09$\\ 
SC\_127 & 110.70747 & -73.49255 & $21.9 \pm 1.2$ & $32.1 \pm 0.8$ & $25.4 \pm 0.5$ & $34.1 \pm 1.2$ & $32.7 \pm 1.1$ & $33.2 \pm 1.5$ & $6.95 \pm 0.02$\\ 
SC\_128 & 110.68775 & -73.48465 & $15.1 \pm 1.2$ & $18.6 \pm 0.8$ & $13.4 \pm 0.7$ & $18.7 \pm 0.5$ & $21.6 \pm 0.8$ & $19.9 \pm 1.5$ & $6.79 \pm 0.03$\\ 
SC\_129 & 110.69481 & -73.48426 & $9.5 \pm 1.2$ & $13.6 \pm 0.7$ & $20.6 \pm 0.7$ & $37.8 \pm 1.2$ & $34.1 \pm 1.0$ & $35.6 \pm 1.5$ & $6.58 \pm 0.06$\\ 
SC\_130 & 110.70230 & -73.48263 & $8.9 \pm 1.2$ & $12.4 \pm 0.8$ & $12.8 \pm 0.7$ & $13.5 \pm 0.8$ & $14.8 \pm 1.1$ & $17.8 \pm 1.5$ & $6.56 \pm 0.06$\\ 
SC\_131 & 110.69502 & -73.48136 & $15.5 \pm 1.2$ & $18.3 \pm 0.7$ & $19.4 \pm 0.7$ & $29.6 \pm 1.2$ & $28.7 \pm 1.1$ & $24.5 \pm 1.5$ & $6.80 \pm 0.03$\\ 
SC\_132 & 110.67261 & -73.47396 & $15.6 \pm 1.2$ & $21.2 \pm 0.8$ & $19.9 \pm 0.7$ & $29.5 \pm 1.2$ & $21.2 \pm 1.1$ & $15.9 \pm 1.5$ & $6.81 \pm 0.03$\\ 
SC\_133 & 110.85435 & -73.45440 & $8.1 \pm 1.2$ & $10.3 \pm 0.6$ & $10.5 \pm 0.7$ & $7.0 \pm 1.2$ & $5.1 \pm 1.1$ & $3.9 \pm 1.5$ & $6.66 \pm 0.06$\\ 
SC\_134 & 110.85244 & -73.45419 & $3.8 \pm 1.2$ & $13.3 \pm 0.6$ & $13.6 \pm 0.7$ & $13.0 \pm 1.2$ & $8.3 \pm 0.6$ & $3.6 \pm 1.5$ & $6.61 \pm 0.15$\\ 
SC\_135 & 110.85889 & -73.45459 & $4.2 \pm 1.2$ & $12.8 \pm 0.8$ & $9.7 \pm 0.6$ & $6.7 \pm 1.2$ & $3.9 \pm 0.8$ & $0.6 \pm 1.5$ & $6.37 \pm 0.10$\\ 
SC\_136 & 110.85791 & -73.45462 & $6.6 \pm 1.2$ & $10.5 \pm 0.7$ & $10.9 \pm 0.6$ & $8.3 \pm 0.7$ & $5.7 \pm 1.1$ & $-1.6 \pm 1.5$ & $6.43 \pm 0.08$\\ 
SC\_137 & 110.85896 & -73.45420 & $6.3 \pm 1.2$ & $11.6 \pm 0.6$ & $12.0 \pm 0.6$ & $10.6 \pm 0.9$ & $4.7 \pm 1.1$ & $5.1 \pm 1.5$ & $6.46 \pm 0.10$\\ 
SC\_138 & 110.86045 & -73.45428 & $10.7 \pm 1.2$ & $12.3 \pm 0.8$ & $13.8 \pm 0.7$ & $12.1 \pm 1.2$ & $7.3 \pm 1.1$ & $6.5 \pm 1.5$ & $6.12 \pm 0.06$\\ 
SC\_139 & 110.86098 & -73.45427 & $6.2 \pm 1.2$ & $5.4 \pm 0.6$ & $6.0 \pm 0.7$ & $6.1 \pm 1.2$ & $4.0 \pm 1.1$ & $-0.0 \pm 1.5$ & $6.43 \pm 0.07$\\ 
SC\_140 & 110.85891 & -73.45394 & $10.7 \pm 1.2$ & $9.3 \pm 0.7$ & $10.6 \pm 0.6$ & $7.4 \pm 0.6$ & $4.3 \pm 1.1$ & $1.0 \pm 1.5$ & $6.15 \pm 0.05$\\ 
SC\_141 & 110.82008 & -73.46628 & $21.7 \pm 1.2$ & $27.3 \pm 0.3$ & $18.4 \pm 0.7$ & $15.3 \pm 0.3$ & $9.1 \pm 0.8$ & $5.6 \pm 1.5$ & $6.17 \pm 0.03$\\ 
SC\_142 & 110.82292 & -73.46619 & $0.9 \pm 1.2$ & $8.3 \pm 0.8$ & $9.5 \pm 0.7$ & $6.9 \pm 1.2$ & $5.4 \pm 1.1$ & $0.4 \pm 1.5$ & $6.29 \pm 0.35$\\ 
SC\_143 & 110.79685 & -73.45404 & $2.9 \pm 1.2$ & $7.2 \pm 0.7$ & $8.4 \pm 0.4$ & $7.2 \pm 1.2$ & $3.7 \pm 0.3$ & $3.5 \pm 1.5$ & $6.08 \pm 0.22$\\ 
SC\_144 & 110.66959 & -73.47230 & $4.1 \pm 1.2$ & $15.2 \pm 0.5$ & $14.1 \pm 0.5$ & $20.5 \pm 0.5$ & $61.5 \pm 1.0$ & $76.3 \pm 1.5$ & $6.21 \pm 0.13$\\ 
SC\_145 & 110.81957 & -73.45875 & $11.0 \pm 1.2$ & $13.2 \pm 0.8$ & $13.6 \pm 0.4$ & $9.7 \pm 1.2$ & $6.1 \pm 0.6$ & $5.3 \pm 1.5$ & $6.64 \pm 0.05$\\ 
SC\_146 & 110.82110 & -73.45871 & $5.1 \pm 1.2$ & $12.6 \pm 0.5$ & $12.4 \pm 0.6$ & $9.3 \pm 1.2$ & $4.7 \pm 1.1$ & $4.1 \pm 1.5$ & $6.37 \pm 0.10$\\ 
SC\_147 & 110.82095 & -73.45843 & $3.9 \pm 1.2$ & $8.1 \pm 0.7$ & $7.4 \pm 0.7$ & $6.5 \pm 1.2$ & $3.2 \pm 1.1$ & $2.5 \pm 1.5$ & $6.33 \pm 0.14$\\ 
SC\_148 & 110.81494 & -73.45791 & $9.8 \pm 1.2$ & $6.5 \pm 0.7$ & $8.5 \pm 0.5$ & $8.7 \pm 0.5$ & $6.0 \pm 1.1$ & $4.2 \pm 1.5$ & $5.93 \pm 0.05$\\ 
SC\_149 & 110.81525 & -73.45789 & $4.7 \pm 1.2$ & $9.2 \pm 0.3$ & $11.1 \pm 0.5$ & $11.4 \pm 0.7$ & $6.3 \pm 1.1$ & $5.7 \pm 1.5$ & $6.22 \pm 0.09$\\ 
SC\_150 & 110.81370 & -73.45776 & $10.4 \pm 1.2$ & $11.2 \pm 0.6$ & $11.8 \pm 0.5$ & $11.3 \pm 0.7$ & $6.9 \pm 1.0$ & $8.0 \pm 1.5$ & $5.96 \pm 0.06$\\ 
SC\_151 & 110.81586 & -73.45765 & $15.9 \pm 1.2$ & $25.2 \pm 0.6$ & $24.4 \pm 0.6$ & $22.4 \pm 1.2$ & $12.1 \pm 0.6$ & $9.4 \pm 0.3$ & $6.95 \pm 0.03$\\ 
SC\_152 & 110.81782 & -73.45708 & $21.2 \pm 1.2$ & $19.1 \pm 0.7$ & $20.1 \pm 0.6$ & $16.1 \pm 1.2$ & $9.0 \pm 1.1$ & $5.7 \pm 1.5$ & $6.46 \pm 0.02$\\ 
SC\_153 & 110.82499 & -73.45730 & $9.6 \pm 1.2$ & $27.0 \pm 0.7$ & $31.1 \pm 0.5$ & $26.5 \pm 0.6$ & $14.6 \pm 1.1$ & $10.0 \pm 0.5$ & $6.85 \pm 0.05$\\ 
SC\_154 & 110.82503 & -73.45740 & $6.7 \pm 1.2$ & $17.6 \pm 0.3$ & $16.3 \pm 0.5$ & $16.4 \pm 1.2$ & $15.2 \pm 1.1$ & $13.1 \pm 1.5$ & $6.78 \pm 0.07$\\ 
SC\_155 & 110.82519 & -73.45764 & $4.6 \pm 1.2$ & $9.7 \pm 0.7$ & $10.7 \pm 0.7$ & $8.3 \pm 1.2$ & $5.0 \pm 0.7$ & $5.9 \pm 1.5$ & $6.50 \pm 0.12$\\ 
SC\_156 & 110.82888 & -73.45774 & $10.3 \pm 1.2$ & $16.0 \pm 0.7$ & $17.5 \pm 0.7$ & $13.6 \pm 1.2$ & $8.8 \pm 1.1$ & $3.5 \pm 1.5$ & $6.36 \pm 0.04$\\ 
SC\_157 & 110.88178 & -73.45305 & $5.6 \pm 1.2$ & $13.1 \pm 0.8$ & $16.2 \pm 0.5$ & $12.4 \pm 1.2$ & $8.2 \pm 1.1$ & $4.5 \pm 1.5$ & $6.63 \pm 0.09$\\ 
SC\_158 & 110.87357 & -73.45284 & $9.0 \pm 1.2$ & $12.3 \pm 0.8$ & $10.6 \pm 0.7$ & $10.0 \pm 1.2$ & $4.2 \pm 1.1$ & $4.9 \pm 1.5$ & $6.52 \pm 0.05$\\ 
SC\_159 & 110.87397 & -73.45293 & $7.5 \pm 0.1$ & $7.0 \pm 0.5$ & $7.6 \pm 0.7$ & $9.3 \pm 1.2$ & $5.0 \pm 1.0$ & $1.6 \pm 1.5$ & $5.72 \pm 0.01$\\ 
SC\_160 & 110.87266 & -73.45103 & $9.4 \pm 1.2$ & $7.5 \pm 0.6$ & $7.1 \pm 0.3$ & $11.3 \pm 1.2$ & $5.6 \pm 1.1$ & $2.6 \pm 1.5$ & $5.82 \pm 0.06$\\ 
SC\_161 & 110.87221 & -73.45115 & $5.2 \pm 1.2$ & $5.5 \pm 0.4$ & $10.9 \pm 0.5$ & $9.1 \pm 1.2$ & $9.1 \pm 1.1$ & $8.8 \pm 0.3$ & $5.97 \pm 0.09$\\ 
SC\_162 & 110.75752 & -73.45268 & $7.9 \pm 1.2$ & $11.5 \pm 0.8$ & $12.1 \pm 0.4$ & $10.6 \pm 1.2$ & $5.3 \pm 1.1$ & $6.3 \pm 1.5$ & $6.26 \pm 0.06$\\ 
SC\_163 & 110.75739 & -73.45370 & $8.8 \pm 1.2$ & $14.7 \pm 0.8$ & $11.3 \pm 0.5$ & $6.9 \pm 1.2$ & $6.8 \pm 1.1$ & $5.5 \pm 1.5$ & $6.52 \pm 0.05$\\ 
SC\_164 & 110.74766 & -73.45285 & $18.4 \pm 1.2$ & $22.9 \pm 0.2$ & $20.9 \pm 0.5$ & $20.5 \pm 1.2$ & $10.0 \pm 1.1$ & $8.1 \pm 1.5$ & $6.83 \pm 0.03$\\ 
SC\_165 & 110.74990 & -73.44874 & $21.2 \pm 1.2$ & $29.8 \pm 0.4$ & $32.3 \pm 0.5$ & $27.7 \pm 0.5$ & $17.5 \pm 0.6$ & $14.4 \pm 1.5$ & $6.68 \pm 0.02$\\ 
SC\_166 & 110.77654 & -73.44677 & $13.1 \pm 1.2$ & $12.5 \pm 0.7$ & $10.7 \pm 0.7$ & $8.0 \pm 0.5$ & $5.9 \pm 0.7$ & $3.1 \pm 1.5$ & $6.12 \pm 0.03$\\ 
SC\_167 & 110.77515 & -73.44679 & $10.0 \pm 1.2$ & $10.5 \pm 0.3$ & $7.4 \pm 0.3$ & $16.1 \pm 1.2$ & $13.3 \pm 1.1$ & $10.1 \pm 1.5$ & $5.84 \pm 0.04$\\ 
SC\_168 & 110.78108 & -73.44683 & $9.3 \pm 1.2$ & $8.7 \pm 0.7$ & $10.2 \pm 0.3$ & $11.1 \pm 1.2$ & $6.4 \pm 1.1$ & $7.3 \pm 0.4$ & $5.98 \pm 0.05$\\ 
SC\_169 & 110.80459 & -73.44832 & $10.5 \pm 1.2$ & $10.6 \pm 0.7$ & $7.2 \pm 0.6$ & $6.7 \pm 1.2$ & $5.2 \pm 0.9$ & $1.5 \pm 1.5$ & $5.90 \pm 0.05$\\ 
SC\_170 & 110.79816 & -73.44769 & $5.8 \pm 1.2$ & $11.9 \pm 0.5$ & $10.0 \pm 0.3$ & $10.9 \pm 1.0$ & $7.9 \pm 0.1$ & $2.7 \pm 1.5$ & $5.79 \pm 0.09$\\ 
SC\_171 & 110.79823 & -73.44778 & $11.9 \pm 1.2$ & $15.5 \pm 0.6$ & $12.8 \pm 0.4$ & $10.3 \pm 1.2$ & $6.8 \pm 1.1$ & $5.0 \pm 1.5$ & $6.21 \pm 0.04$\\ 
SC\_172 & 110.86482 & -73.45783 & $3.7 \pm 1.2$ & $13.8 \pm 0.6$ & $16.7 \pm 0.6$ & $13.5 \pm 1.2$ & $6.4 \pm 1.1$ & $3.0 \pm 1.5$ & $6.65 \pm 0.14$\\ 
SC\_173 & 110.86402 & -73.45812 & $6.7 \pm 1.2$ & $10.8 \pm 0.7$ & $10.2 \pm 0.5$ & $10.1 \pm 1.2$ & $6.6 \pm 0.6$ & $6.9 \pm 1.5$ & $5.77 \pm 0.09$\\ 
SC\_174 & 110.90066 & -73.45376 & $7.9 \pm 1.2$ & $13.4 \pm 0.7$ & $12.1 \pm 0.6$ & $10.2 \pm 1.0$ & $5.4 \pm 1.1$ & $3.4 \pm 0.4$ & $6.62 \pm 0.07$\\ 
SC\_175 & 110.70823 & -73.47715 & $32.5 \pm 1.2$ & $23.8 \pm 0.5$ & $17.1 \pm 0.5$ & $19.8 \pm 1.2$ & $25.9 \pm 1.1$ & $18.0 \pm 1.5$ & $7.12 \pm 0.01$\\ 
SC\_176 & 110.68084 & -73.47226 & $18.9 \pm 1.2$ & $24.8 \pm 0.6$ & $18.8 \pm 0.5$ & $12.3 \pm 0.6$ & $10.2 \pm 1.1$ & $7.9 \pm 1.5$ & $6.89 \pm 0.03$\\ 
SC\_177 & 110.64783 & -73.47067 & $5.3 \pm 1.2$ & $10.4 \pm 0.8$ & $10.6 \pm 0.6$ & $14.5 \pm 1.2$ & $12.2 \pm 1.1$ & $8.8 \pm 1.5$ & $6.33 \pm 0.09$\\ 
\end{longtable*}
\end{center}

\newpage
\begin{deluxetable*}{l l l l l l}
\tabletypesize{\scriptsize}
\tablecaption{Stacked photometry for all of the compact star clusters. All flux densities have been corrected for Galactic extinction.\label{tab:photostack}}
\tablewidth{0pt}
\tablehead{
\colhead{F090W} & \colhead{F150W} & \colhead{F200W} & \colhead{F277W} & \colhead{F356W} & \colhead{F444W}\\[-0.2cm]
\colhead{(nJy)} & \colhead{(nJy)} & \colhead{(nJy)} & \colhead{(nJy)} & \colhead{(nJy)} & \colhead{(nJy)}
}
\startdata
$9.79 \pm 0.12$ & $13.11 \pm 0.07$ & $13.98 \pm 0.06$ & $13.02 \pm 0.12$ & $8.16 \pm 0.11$ & $6.64 \pm 0.15$
\enddata
\end{deluxetable*}


\bibliography{bibli}{}
\bibliographystyle{aasjournal}



\end{document}